\documentclass[12pt]{article}
\pdfoutput=1
\usepackage{amsmath}
\usepackage{amssymb}
\usepackage{graphicx,bbm,mathrsfs}
\usepackage{nicefrac}
\usepackage{bbm}
\usepackage{geometry}

\usepackage{mathtools}

\usepackage{dcolumn}   
\usepackage{bm}        
\usepackage{graphicx,mathrsfs}
\usepackage{nicefrac}
\usepackage{multirow}
\usepackage{color}
\usepackage{relsize}
\usepackage{adjustbox}

\newcommand{\be}{\begin{equation}}
\newcommand{\ee}{\end{equation}}
\newcommand{\bea}{\begin{eqnarray}}
\newcommand{\eea}{\end{eqnarray}}

\usepackage{wasysym}

\begin{document}

\vspace*{1.2cm}

\begin{center}

\thispagestyle{empty}
{\Large
Braneworld Effective Field Theories--\\ \vspace{10pt}
Holography,  Consistency and Conformal Effects
 }\\[10mm]

\renewcommand{\thefootnote}{\fnsymbol{footnote}}

{  Sylvain~Fichet
\footnote{sfichet@caltech.edu } }\\[10mm]

\end{center} 
\noindent
\quad\quad\textit{Walter Burke Institute for Theoretical Physics, California Institute of Technology,}

\noindent \quad\quad \textit{Pasadena, CA
91125, California, USA} \\

\noindent
\quad\quad \textit{ICTP South American Institute for Fundamental Research  \& IFT-UNESP,}

\noindent \quad\quad \textit{R. Dr. Bento Teobaldo Ferraz 271, S\~ao Paulo, Brazil
}

\addtocounter{footnote}{-1}

\vspace*{12mm}

\begin{center}
{  \bf  Abstract }
\end{center}

Braneworld theories are often described as  low-energy effective field theories  (EFTs) featuring an infinitely thin $3$-brane and 4D fields exactly localized on it. 
We investigate whether an exactly localized braneworld can arise as a limit of a theory  of 5D  fields.
Using a holographic formalism we argue that such limit does not exist in the presence of gravity, therefore implying  a discontinuity in the space of EFTs. 
We then present specific models involving exactly localized fields in which inconsistencies appear,   which are solved when  fields are taken as quasilocalized. Part of our arguments  rely on conjectures from the ``swampland'' program.

Our investigation motivates  braneworld EFTs built from 5D fields, \textit{i.e.} quasilocalized braneworlds.
Observable effects from quasilocalization are significant for warped braneworlds such as Randall-Sundrum~II (RSII),
and are reminiscent of 
a conformal hidden sector. 
Focusing on the gauge-gravity sector we show that mani\-festations of the quasilocalized warped braneworld include \textit{i)} an anomalous running of SM gauge couplings \textit{ii)} a  conformal contribution to SM gauge boson scattering induced by 5D gravity. 
Constraining  these effects puts an upper bound on the 5D EFT cutoff, implying that
the warped braneworld hypothesis could---at least in principle---be tested completely.

\newpage
\tableofcontents
\newpage

\section{Introduction}

In quantum field theory, branes are space-filling hypersurfaces located in a higher-dimensional spacetime. 
Branes may be viewed as solitons on which particles  can be localized. 
Similar objects naturally appear in string theory  as D-branes, which are dynamical objects with quantum properties \cite{Polchinski:1996na,Bachas:1998rg}. Black brane solutions also arise in the supergravity limit of string theories \cite{Aharony:1999ti, Duff:1996zn}. 

From the effective  field theory viewpoint, branes are simply described as infinitely thin surfaces being part of the  background in the fundamental action of the theory \cite{Csaki:2004ay, Sundrum:1998sj,Sundrum:1998ns}. 
A brane can have matter fields localized on it, a feature at the center of our attention in this work. Here we will generically refer to any theory with such brane-localized matter as a  ``braneworld''.

In effective field theory, there is no principle forbidding matter fields to live exclusively in the worldvolume of a brane. 
This kind of EFT has been used in  early proposals of braneworld models (\textit{e.g.} \cite{Dvali:2000hr,Randall:1999ee,Randall:1999vf}). 
In contrast, it is also possible to write Lagrangians where some operators are localized on the brane, while the matter fields themselves live in the entire spacetime. In this case, certain degrees of freedom encoded in the higher-dimensional fields can still be localized towards the brane, without being strictly confined on it. 
We  therefore have two kinds of theories, here referred to as ``exactly localized'' and ``quasilocalized'' braneworlds.

The distinction between these two kinds of braneworld EFTs might seem at first view somewhat artificial. It may seem reasonable to expect that an exactly localized braneworld can simply arise as a limit of a quasilocalized braneworld. 
However, we will see that such equivalence is in general not true and that the situation is in reality more subtle. 
  This is the starting observation made in this work. It will then lead us to reconsider consistency of exactly localized braneworlds and to study observable effects  from quasilocalized braneworlds.

In Secs.~\ref{se:EFT}--\ref{se:limit}, we introduce the formalism and  make clear that the infinite localization limit can come from either bulk masses or brane kinetic terms. 
To consistently compare exactly and quasi-localized theories, the quasilocalized braneworld is treated via a holographic formalism---in which variables are exactly brane-localized. We then show in Sec.~\ref{se:disc}  that, at the very least in the presence of gravity, exactly localized braneworlds do not arise as a limit of quasilocalized ones.

The existence of a discontinuity in theory space
leads us to further scrutinize  exactly localized braneworlds.
 In Sec.\,\ref{se:swamp}, considering simple, specific models with exactly localized fields, we find that inconsistencies arise in the presence of gravity. 
Some of the arguments rely on standard conjectures from the swampland program. 

The discontinuity between the two kinds of braneworld EFTs and the hints of inconsistency of the (field theoretical) exactly localized braneworld  naturally lead us to revisit braneworld models which were initially proposed as exactly localized. 
 As a general feature, quasilocalized braneworlds have a richer phenomenology than exactly localized ones. 
In Sec.~\ref{se:RS}, we focus on a quasilocalized version of the Randall-Sundrum II model. 
While the original model only has 5D gravity, the quasilocalized model has a whole matter sector in the bulk, naturally behaving as a conformal hidden sector---this property has recently inspired warped dark sector model-building \cite{Brax:2019koq,Costantino:2019ixl}. 
Focusing on the gauge-gravity sector, which is especially model-independent, we present two physical effects implied by gauge field quasilocalization---which are absent in the exactly localized version of the warped braneworld.

\section{Braneworld effective theories}
\label{se:EFT}

Our focus is on codimension-1 branes \textit{i.e.}  branes that span one dimension less than the dimension of the full spacetime. For convenience, and although it is not mandatory for most of the conceptual discussions in the paper, we shall restrict to  4+1 spacetime and therefore focus on 3-branes.

We are interested in  3-branes that are Poincar\'e invariant. We write thus a general five-dimensional metric
\be
ds^2=g_{MN}dX^M dX^N = e^{-2a(y)}\eta_{\mu\nu}dx^\mu dx^\nu - dy^2  \,, \label{eq:metric}
\ee
where  $a(y)=0$ corresponds to the flat extradimension case and  $a(y)=ky$ corresponds to $AdS_5$ space with curvature $k$. $\eta_{\mu\nu}$ is Minkowski metric with signature $(+,-,-,-)$.

As customary in higher dimensional EFTs,  we model a 3-brane as an infinitely thin surface. Comments  on that aspect will be made in Sec.~\ref{se:width}.
The brane is centered on the position $y=y_0$ of the extradimension. 
In our discussion we will sometimes assume the existence of a second brane at $y=y_1\equiv y_0+L$. This second brane can be removed from the theory by taking $L\rightarrow \infty$.

\subsection{Localized and quasilocalized EFTs}

When defining a braneworld effective theory,  it is commonplace  to allow matter fields exactly localized on the brane, 
\be
\tilde S=S_5+ \int d^4x \sqrt{|\bar g|} \left( {\cal L}\left[\phi,\psi,A^\mu\right] +\ldots \right)\bigg|_{y=y_0}  \, \label{eq:Stilde}
\ee
where  the $\phi,\psi,A^\mu$ fields are function of $x^\mu$ only, and the 5D component of the action $S_5$ is independent of these fields. 
Including such exactly localized degrees of freedom is compatible with all the symmetries left unbroken on the brane. 
In Eq.~\eqref{eq:Stilde}, $\bar g_{\mu\nu}$ is the induced metric on the brane. 
The ellipses correspond to brane-localized operators independent  of the matter fields, such as a brane-localized Ricci scalar, a brane tension, and the Gibbons-Hawking-York term.
We  refer to the EFT in Eq.~\eqref{eq:Stilde} as an \textit{exactly localized} braneworld. We will use a tilde superscript to denote quantities associated with this kind of EFT.

It is also possible to write a different kind of  braneworld effective theory where a set of \textit{operators} is localized on the brane,  while all matter fields of the theory are five dimensional.  
The action in that case reads 
\be
S= S_5\left[\Phi,\Psi, {\cal A}^M\right] + \int  d^4x \left( \sqrt{|\bar g|}  {\cal L}_4\left[\Phi,\Psi, {\cal A}^M\right]  +\ldots \right)\bigg|_{y=y_0}\, \label{eq:Sgen}
\ee
where the brane operators are encoded in ${\cal L}_4$ and the 5D fields $\Phi,\Psi, {\cal A}^M$ depend on $X^M$. The 5D  action $S_5$ depends on the 5D fields and contains operators such as the 5D kinetic terms
\be
S_\Phi^{\rm kin}= \int d^5X \sqrt{g}\left(
\frac{1}{2}\partial_M\Phi \partial^M \Phi -\frac{1}{2}m^2_\Phi \Phi^2 \right)   \label{eq:5Dscal_kin}
\ee
\be
S_\Psi^{\rm kin} = \int d^5X \sqrt{g}\left(
\frac{i}{2}\left(
\bar \Psi \Gamma^M D_M \Psi- D_M\bar \Psi \Gamma^M  \Psi
\right) -m_\Psi \bar \Psi \Psi \right)  
\label{eq:5Dferm_kin}
\ee
\be
S_A^{\rm kin} = \int d^5X \sqrt{g}\left(
-\frac{1}{4g^2_5} {\cal F}^{MN}{\cal F}_{MN} \right)   \,.
\label{eq:5Dgauge_kin}
\ee

In this second type of EFT, for appropriate choices of parameters of the 5D and brane Lagrangians in Eq.~\eqref{eq:Sgen}, a  degree of freedom with  4D properties can exist in the spectrum and be  almost localized on the brane. This feature will be studied in details in Sec.~\ref{se:limit}. 
Such highly localized limit of the theory defined in Eq.~\eqref{eq:Sgen} is the central focus of this work.
With such limit in mind, we will be readily refering to the EFT in Eq.~\eqref{eq:Sgen} as a \textit{quasilocalized} braneworld.

One could of course write theories mixing both 5D fields and  exactly localized 4D fields. It turns out that this mixed case does not require dedicated discussion, hence  no naming is needed. Only in Sec.~\ref{se:swamp} a model of this kind will be studied. In the rest of the paper it is enough to consider actions where  matter fields are either all exactly localized or all quasilocalized 5D fields.

It is natural to ask how the two kinds of EFT defined above---the exactly and quasilocalized braneworlds---relate to each other. 
 Can the exactly localized braneworld arise as a limit of the quasilocalized braneworld?

This is the central question we want to address in Secs.~\ref{se:limit},\,\ref{se:disc}. The  proper way to define the question is to compare the physical observables of both theories, and therefore to compare their correlation functions. We will thus work at the level of the quantum effective actions.

\subsection{Quantum actions  and braneworld holography}

In this section we only consider scalar fields. The formalism for other spins is essentially similar although more technical. We work at the level of the quantum effective action $\Gamma$,  which encodes all information about correlation functions. To avoid naming confusion, we refer to $\Gamma$  as the ``quantum action''. 

For the exactly localized braneworld theory, the  quantum action is given by\,\footnote{ The 1PI index indicates that only 1PI diagrams are selected in the path integral. This is a shortcut notation for the usual construction of the generating functionals,
\be 
Z[J]=\int{\cal D} \phi \exp\left(i \tilde S[\phi] + i\int d^4x \phi J \right)=\exp\left(i W[J]\right) \,, \quad \Gamma[\phi_{\rm cl}]=W[J]- \int d^4x \phi_{\rm cl} J 
\label{se:Gamma_def}\,.
\ee
The argument of $\Gamma$ is always a classical field value. The ``$\rm cl$'' index will not be specified throughout the text. 
}
\be
\exp\left( i \tilde \Gamma[\phi] \right) = \int_{\rm 1PI} {\cal D} \hat\phi \exp\left( i \tilde S[\phi+\hat \phi] \right) \,. \label{eq:Gamtilde}
\ee
Spacetime has five dimensions, and interacting 5D theories  always are low-energies EFTs. 
 The predictions arising from $\tilde \Gamma[\phi]$ are  only valid  up to an energy scale of order $\tilde \Lambda$ (or a distance scale of order $1/\tilde \Lambda$), the validity cutoff of the EFT. Beyond this scale the theory should be superseded with a UV-completion.

Let us turn to the quantum action for the quasilocalized braneworld. Since we aim to study  the  quantum action of the quasilocalized braneworld in a limit potentially reproducing  $\tilde \Gamma[\phi]$, the  quantum action  should be expressed in terms of a classical variable that can  match the exactly localized  variable $\phi$ of $\tilde \Gamma[\phi]$  in the limit of infinite localization.  A logical choice is to express the quantum action of the quasilocalized braneworld as a function of the classical value of the 5D field on the brane, $\Phi_0\equiv \Phi(y=y_0) $. 
This is the definition of  a \textit{holographic} formalism, where $\Phi_0$ is the holographic variable (see \textit{e.g.} \cite{Aharony:1999ti,   Nastase:2007kj, Gherghetta:2010cj, Ponton:2012bi, Witten:1998qj}).

From now on we work in momentum space for the $x^\mu$ coordinates, introducing $\Phi( p_\mu,z)=\int d^4x e^{ix^\mu p_\mu } \Phi(X^M)$. One also defines the absolute momentum $p=\sqrt{\eta_{\mu\nu}  p^\mu p^\nu }$, which is real (imaginary) for timelike (spacelike) momentum. 
The 5D field in position-momentum space, $\Phi(p^\mu,y)$, is rewritten as
\be
\Phi(p^\mu,y)= \Phi_0(p^\mu) K(p^\mu,y)\,, \quad \quad {\rm with} \quad K(p^\mu,y_0)=1\,. \label{eq:Phi0def}
\ee
The meaning of $K$ will become obvious in the semi-classical expansion detailed in next section.

Using Eq.~\eqref{eq:Phi0def} in the definition of the action, the  quasilocalized braneworld is  described by the holographic quantum action
\be
\exp\left( i \Gamma[\Phi_0] \right) = \int_{\rm 1PI} {\cal D} \hat\Phi \exp\left( i S[\Phi_0 K+\hat \Phi] \right) \,. \label{eq:Gam}
\ee
As for the exactly localized case, since the theory is five-dimensional the correlators are valid up to a UV  cutoff denoted $\Lambda$. 

With these definitions, the question of exact localization can be formally expressed using $\Gamma, \tilde \Gamma$ and thinking in terms of parameter space. What we ask is whether there exists a direction in the parameter space  of the quasilocalized  
braneworld Lagrangian (Eq.~\eqref{eq:Sgen}) for which 
\be\Gamma \rightarrow \tilde \Gamma \,. \ee 
This question will be addressed in Secs.~\ref{se:limit} and \ref{se:disc}.

Finally we emphasize that the holographic formalism   introduced above can be introduced for any boundary and any metric, and has thus in itself nothing to do with the AdS/CFT duality.\,\footnote{The AdS/CFT aspect appears when  the 5D metric is  AdS$_5$, at least asymptotically near the UV brane.
}

\section{Holographic  action and the exact localization limit}
\label{se:limit}

In the validity regime of the 5D EFT, the 5D interactions (including gravity) can be treated perturbatively. We can thus expand and truncate the quasilocalized braneworld action in powers of $\hbar$, \textit{i.e.} in the  semiclassical expansion, such that
\be
\Gamma[\Phi_0]= \Gamma_{\rm cl}[\Phi_0] + \ldots \label{eq:Gammaexp}
\ee
Here $\Gamma_{\rm cl}$ is the classical holographic action and  the ellipses represent the 1-loop functional determinant and higher order Feynman diagrams. 
The classical bulk field $\Phi(p^\mu,y)$ satisfies the classical 5D equation of motion (EOM), and has fixed value $\Phi_0(p^\mu)$ on the brane.

In order to determine the content of the holographic action, we will need the Feynman propagator of $\Phi$ with Neumann boundary condition on the brane. 
This Neumann propagator in position-momentum space $(p_\mu,y)$ is denoted
\be
\langle \Phi( p^\mu,y)  \Phi( -p^\mu,y')\rangle \equiv \Delta_{p}(y,y')\equiv i G_{p}(y,y')\,. 
\ee
A derivation of the general Feynman propagator in the conformally flat background of Eq.~\eqref{eq:metric} is given in App.~\ref{se:propa_gen}. 

Let us now consider the holographic profile $K(p^\mu,y)\equiv K_p(y)$ from Eq.~\eqref{eq:Phi0def} in the classical regime. The classical $K_p(y)$ satisfies the 5D  EOM, $K_p(y_0)=1$ and another boundary or regularity condition that the Neumann propagator satisfies as well. Since the propagator has the structure $\Delta_{p}(y,y')\propto F_<(y_<)F_>(y_>)$  where $y_<=\min(y,y')$, $y_>=\max(y,y')$ and the $F$ functions satisfy the homogeneous 5D EOM, it follows that 
 \be
 K_p(y) =  \frac{G_{p}(y_0,y)}{G_{p}(y_0,y_0)}\,. \label{eq:K_class}
 \ee
This  relation can be explicitly checked using the general form of the propagator in Eq.~\eqref{eq:propa_gen}. 
 In other words,  the classical profile is equal to the ``amputated brane-to-bulk propagator''.\,\footnote{``Amputation'' refers to the removal of  $G_{p}(y_0,y_0)$.  }

Let us now consider the bilinear part of the  holographic action, which contains information on the spectrum of the theory. It reads
\be
\Gamma_{\rm cl}[\Phi_0]= \frac{1}{2}\int \frac{d^4p}{(2\pi)^4} \int dy e^{-4 a(y)} \left( e^{2a(y)} p^2 \Phi^2 -(\partial_5\Phi)^2-m^2_\Phi \Phi^2
+ \delta(y-y_0) {\cal L}_4'' \Phi^2
  \right)+ \ldots \label{eq:GamPhi0}
\ee
with 
${\cal L}_4''=\frac{\delta^2}{\delta \Phi \delta \Phi}{\cal L}_4\big|_{\Phi=0}$. Integrating Eq.~\eqref{eq:GamPhi0} by part makes appear the brane operator ${\cal B} \Phi=\partial_5 \Phi(y_0)+ {\cal L}''_4 \Phi(y_0) $ and the classical 5D EOM. The EOM piece vanishes and the non-vanishing part of the bilinear action comes from the remaining boundary terms,
\be
\Gamma[\Phi_0]= \frac{1}{2} \int \frac{d^4p}{(2\pi)^4} 
\Phi_0(p) \Pi_\Phi(p) \Phi_0(-p)
+ \ldots 
\ee
where \be
\Pi_\Phi(p)\equiv {\cal B}K_p(y)\, 
\ee
is the ``holographic self-energy'' and ${\cal B}$ is the boundary operator 
(see App.~\ref{se:propa_gen}). 
Evaluating ${\cal B}K_p(y)$ using the explicit expression of the propagator in Eq.~\eqref{eq:propa_gen}, one finds that the holographic self-energy is given by the inverse of the brane-to-brane propagator, 
\be
\Pi_\Phi(p) = \frac{1}{G_p(y_0,y_0)}\,. \label{eq:Pi_def}
\ee
This defines the bilinear piece of the classical holographic action. Let us briefly discuss  the structure of the rest of the action. 

Regarding  interaction terms, the classical holographic action involves spatial overlaps of the holographic profiles from bulk interactions. A $\lambda \Phi^4$ bulk interaction, for instance, when put in position-momentum space, becomes
\be
\delta^{(4)}\left(\sum p_i^\mu\right)\,\lambda\,\Phi_0(p_1^\mu)\Phi_0(p_2^\mu)\Phi_0(p_3^\mu)\Phi_0(p_4^\mu) \int dy K_{p_1}(y)K_{p_2}(y)K_{p_3}(y)K_{p_4}(y)\,. \label{eq:Phi4hol}
\ee
where $p_{1\ldots 4}$ are the absolute four-momentum of the four $\Phi$ fields. 

Finally, the  quantum terms of the holographic action (\textit{i.e.} the higher order terms in Eq.~\eqref{eq:Gammaexp}) encode loops involving the propagator with arbitrary endpoints in the bulk. The endpoints end on   position-momentum space vertices and  are always integrated over the whole bulk.

\subsection{Propagator from brane dressing}

The previous results are fairly standard. 
To further understand the structure of the holographic action and of  the subsequent correlation functions, let us examine how the brane Lagrangian influences the propagator. 
While this seems at first view a nontrivial task, the structure becomes manifest once we choose an appropriate formulation. 

Be $\hat \Delta_p(y,y')$ the Feynman propagator with Neumann boundary condition on the brane and \textit{no} brane Lagrangian, \textit{i.e.} 
\be
\hat \Delta_p(y,y') \equiv \hat \Delta_p(y,y') \bigg|_{{\cal L}_4=0} \,.
\ee
Let us then use the identity
\begin{align}
 \hat \Delta_p(y,y') & =  \frac{\hat \Delta_p(y_0,y) \nonumber \hat \Delta_p(y_0,y')}{\hat \Delta_p(y_0,y_0)}+\hat \Delta^{D}_p(y,y') \\ & =  
  i  \frac{\hat K_p(y)  \hat K_p(y')}{\hat \Pi_p(y_0,y_0)}+\hat \Delta^{D}_p(y,y') \,
  \label{eq:hatDelta }
\end{align}
where $\hat \Delta^{D}_p(y,y')$ is the propagator with Dirichlet boundary condition on the brane. In the last line we have introduced the holographic profile and self-energy using relations Eqs.~\eqref{eq:K_class} and ~\eqref{eq:Pi_def}. These are profiles and self-energies defined from $\hat \Delta_p(y,y')$, \textit{i.e.} in the absence of the brane Lagrangian.

To obtain the (exact) propagator in the presence of the brane Lagrangian ${\cal L}_4$, we can dress the $ \hat \Delta_p(y,y')$ propagator with a generic brane localized insertion $-i\kappa(p)\delta(y-y_0)$, with $\kappa(p)=-{\cal L}''_4(p)$ for tree-level insertions. The brane localized insertion  can encode a tree-level effect such as a brane mass or kinetic term, or even a loop diagram induced by brane-localized interactions. The geometric series representation of the propagator in the presence of the brane Lagrangian  reads
\begin{align}
  \Delta_p(y,y') & =   \hat \Delta_p(y,y') -  \hat \Delta_p(y,y_0)i \kappa(p) \hat \Delta_p(y_0,y_0)+  \hat \Delta_p(y,y_0)i \kappa(p) \hat \Delta_p(y_0,y_0)i \kappa(p) \hat \Delta_p(y_0,y')+\ldots  \label{eq:dressing1}
\end{align}

At that point, we can notice explicitly from Eq.~\eqref{eq:dressing1} that the Dirichlet propagator is insensitive to the brane dressing. This implies 
\be\hat \Delta^D_p(y_0,y')=\Delta^D_p(y_0,y')\,. \label{eq:DeltaDsimp}\ee Another, less obvious feature is that the holographic profile itself is independent of the brane dressing, such that
\be
\hat K_p(y)=K_p(y)\,. \label{eq:Ksimp}
\ee
This can be seen using the explicit expressions in App.~\ref{se:propa_gen}, and it can also be deduced by inspecting the result of the summation of Eq.~\eqref{eq:dressing1}.
Taking into account Eqs.~\eqref{eq:DeltaDsimp}, \eqref{eq:Ksimp}, the dressed propagator takes the form 
\be
 \Delta_p(y,y')  =  i\frac{ K_p(y)  K_p(y')}{\hat \Pi(p)-\kappa(p)} +   \Delta^D_p(y,y') \,.
 \label{eq:dressing2}
\ee
This exact expression for the propagator is valid for any metric and spectrum and any kind of brane insertion. It 
is rather enlightening and will be extensively used in the following sections to elucidates properties of the quasilocalized action. 

We can already notice that the expression Eq.~\eqref{eq:dressing2} shows explicitly  that the brane dressing only affects the holographic self-energy. From Eq.~\eqref{eq:dressing2} it follows that the holographic self-energy in the presence of the brane Lagrangian is given by
\be
\Pi(p)= \hat \Pi(p)-\kappa(p)\,. \label{eq:self}
\ee

We can also notice that the Dirichlet contribution in Eq.~\eqref{eq:dressing2} encodes effects which do not appear in the classical piece of the holographic action. The Dirichlet part of the propagator appears only in internal lines, and will thus contribute to quantum parts of the holographic action. The Dirichlet piece will also appear in one-particle reducible diagrams. 

 Finally one may recall that in the ``compositeness'' language,  the Dirichlet modes are understood as purely composite states, \textit{i.e.} states with no mixing with the elementary probe field. However, since our approach is valid for arbitrary metric, it makes clear that the structure of the propagator Eq.~\eqref{eq:dressing2} has in itself nothing to do with  the elementary/composite picture or AdS/CFT. 
 In the context of compositeness, one may also notice
that the form Eq.~\eqref{eq:dressing2} is somewhat reminiscent of the ``holographic basis'' proposed in \cite{Batell:2007jv}:
In both approaches the subset of Dirichlet modes  is made manifest. However it seems there is no simple connexion between the two formalisms.

\subsection{Localization limits }

We now investigate possible exact localization limit(s) of the quasilocalized braneworld. Here we merely identify potential directions in the parameter space, these directions will be further analyzed in Sec.~\ref{se:disc}.

A first necessary condition for realizing $\Gamma\rightarrow \tilde \Gamma$ appears at the level of the spectrum. Since  the exactly localized action describes a 4-dimensional degree of freedom, the holographic self-energy of the quasilocalized action should reproduce a 4D degree of freedom
\be
\Pi(p) \propto p^2-m_0^2 \label{eq:Pilim}
\ee
in the  limit of exact localization.
This condition implies that a 4D mode has to emerge in the spectrum of the quasilocalized theory, and that the rest of the spectrum  has to vanish from the theory in some fashion  in the exactly localized limit. 

A second necessary condition appears at the level of interactions.  When taking the exact localization limit, interactions have  to reduce in some way to the ones of a 4D brane-localized Lagrangian.

\subsubsection{Large bulk mass}
\label{se:BMdir}

For scalar and fermions, a potential direction for exact localization may exist in the limit of large bulk mass $m_\Phi$, $m_\Psi$ (see Eqs.~\eqref{eq:5Dscal_kin},\,\eqref{eq:5Dferm_kin}). 

Let us show that a potential candidate for a 4D mode exists for any metric. 
We  consider a discrete spectrum. For any  metric this can always  be obtained by assuming the presence of a second  brane at finite proper distance---keeping open the possibility of sending this brane to infinity later in the calculation. 

For a discrete spectrum the candidate for the quasilocalized mode is easily identified and exists for any metric. 
This is a mode with approximate exponential profile which is always a solution of the 5D equation of motion. 
The brane Lagrangians ${\cal L}_4$ (see Eq.~\eqref{eq:Sgen}) can be suitably tuned such that this special mode is always present in the spectrum. When massless or light, such mode is usually dubbed ``zero mode''. However this mode can also be very massive---and potentially still exponentially localized, as we will see below. Hence we refer to it as the \textit{special} mode.

The relevant brane where the special mode is localized is here taken to be at $y_0=0$, the second brane is at finite distance $y_1>y_0$. 
The profiles are controlled by the bulk mass parameters $m_\Phi$, $m_\Psi$. The mass of the special mode is controlled by the brane Lagrangians. This can be seen directly in Eq.~\eqref{eq:dressing2}, a brane-localized mass term $\kappa(p)\equiv m^2_b$ directly contributes to the mass of a 4D state $\hat \Pi \sim p^2-m^2_b +\ldots $ and can be used to tune the effective 4D mass of that state.  Whenever this physical 4D mass of the special mode is small with respect to these 5D masses, it is negligible in the equation of motion and thus has  negligible impact on the special mode profile. 
Moreover, the limit of quasilocalization  is the limit of very large bulk mass, which thus allows high mass for the special mode.

Let us consider the kinetic terms of these modes. 
The 5D action for the fermion special modes takes the form
\be
  \int d^4x dy  N_\Psi e^{-a(y)-2|m_\Psi|y} \bar \psi_0 i \gamma_\mu \partial^\mu \psi_0 \,.
\ee
For scalar modes, the equation of motion does not have an analytic solution for arbitrary metric. However in the limit of large bulk mass---which is our focus, the effects of the curved metric can be neglected. The same is true for the fermion and in the large bulk mass limit 
the 5D actions of the special modes 
are approximately
\be
  \int d^4x dy \, 2m_\Phi e^{-2|m_\Phi|y} \partial_\mu \phi_0  \partial^\mu \phi_0 +\ldots \,,\quad   \int d^4x dy \, 2m_\Psi e^{-2|m_\Psi|y} \bar \psi_0 i \gamma_\mu \partial^\mu \psi_0 +\ldots \label{eq:zero_modes}
\ee
where we have now neglected the $a(y)$ terms in the exponential since we are assuming $a(y)\ll m_{\Phi} y, m_{\Psi} y$ near the brane.\,\footnote{
Away from the brane it is possible that $a(y)$ blow up (see \textit{e.g.} \cite{Cabrer:2009we}) such that it is not negligible with respect to the bulk mass term. However in such region the special mode profile is highly suppressed, hence such effect is negligible for our purposes. The approximate profiles in Eq.~\eqref{eq:zero_modes} are essentially set by their behaviour near the brane. }
The above profiles are valid as long as the  4D mass of the special mode  is small with respect to $m_{\Phi}$, $m_{\Psi}$.

So far we have considered a discrete spectrum, possibly enforced by a second brane at $y_1$. 
We now remove this brane, $y_1 \rightarrow \infty$, such that the spectrum may become continuous. 
The kinetic normalization of the modes in Eq.~\eqref{eq:zero_modes} remains finite for $y_1 \rightarrow \infty$  \textit{i.e.} the modes are normalizable. Therefore the existence of the special modes is guaranteed for any metric. 

If the spectrum is continuous,  one subtlety is that the special mode may in principle mix with the KK continuum. 
When it is the case,  such effect would have to vanish in the localization limit for a pure 4D mode to be recovered \textit{i.e.} for Eq.~\eqref{eq:Pilim} to be asymptotically satisfied.

Our analysis here is about the existence of a 4D degree of freedom potentially reproducing the exactly localized limit. However it does not say anything about the rest of the spectrum, and it is thus not clear if the necessary condition Eq.~\eqref{eq:Pilim} is satisfied in the exactly localized limit. It is, as a matter of fact,  a very model-dependent feature, as can be seen by inspecting the KK spectrum of flat and warped cases.
The aspect of interactions of the brane-localized modes  will be treated in more details in Sec.~\ref{se:disc}.

\subsubsection{Large brane kinetic terms }
\label{se:BKTdir}

For fields of any spin, another  limit giving potentially rise to an exactly localized braneworld is the one of  large brane-localized kinetic term. 
The action takes schematically the form
\be
S=\int d^4x dy \left(\sqrt{g}  {\cal L}^{\rm kin}_5 + \delta(y-y_0)\,r \sqrt{\bar g}  {\cal L}^{\rm kin}_4\right)\,
\ee
where  $r$ controls the magnitude of the brane-localized kinetic term.\,\footnote{$r$ has dimension of length.}  Sending $r$ to infinity, one might expect to obtain an exactly localized theory. 

To show that a 4D mode exists at large $r$ for any metric, it is enough to consider the self-energy Eq.~\eqref{eq:dressing2}. The brane kinetic term contributes as $\kappa(p)=-r p^2$ in the propagator. 
For $r\rightarrow \infty$, the brane term overwhelms the bulk term $\hat \Pi(p)$ such that
\be
\Pi(p) \approx r p^2\,. \label{eq:Pi_kin}
\ee
Hence in that limit the holographic action indeed contains  a single 4D mode. 
While this is shown here for the scalar propagator,  the mechanism  is similar for fermion and  gauge fields. 
The limit Eq.~\eqref{eq:Pi_kin} applies for any spectrum, discrete or continuous, and for any metric. Hence 
  Eq.~\eqref{eq:Pi_kin} always ensures that an exact 4D mode  arises asymptotically in the $r\rightarrow \infty$  limit.


The aspect of interactions  will be treated in more details in Sec.~\ref{se:disc}.

\section{Discontinuities in theory space}
\label{se:disc}

In the previous section we have identified potential directions in the parameter space of the quasilocalized braneworld EFT  for which the theory seems to tend to an exactly localized braneworld.  In this section we show that, at least in  the presence of gravity, none of these limits actually lead to an exactly localized braneworld.

\subsection{Obstruction on bulk masses \label{se:mass}}

We have seen that  a localized 4D mode of scalar or fermion can appear if one takes the bulk masses $m_\Phi$, $m_\Psi$  to infinity (see Sec.~\ref{se:BMdir}). 
However,   whenever a 5D theory is interacting, it is a low-energy EFT  with a finite cutoff $\Lambda$. 
Hence, when the theory has 5D interactions, the bulk masses $m_\Phi$, $m_\Psi$ should not exceed $\Lambda$ for the theory to remain valid.\,\footnote{This upper bound on bulk masses has a CFT equivalent as an upper bound on the conformal dimension of CFT operators, see \cite{Fitzpatrick:2010zm}. }
 Therefore bulk masses  cannot go to infinity; there is an obstruction.

One could in principle set the 5D matter interactions of $\Phi$, $\Psi$ to zero, and set all 5D higher-dimensional operators to zero at a given scale. Doing this removes the 5D cutoff in the absence of gravity. However, whenever gravity is present, the 5D theory is interacting hence a finite Planckian cutoff always exists.

Although the above argument is in principle sufficient to discard the possibility of $m_{\Phi, \Psi}\rightarrow \infty$, let us  ignore it and allow arbitrary values of bulk masses. Consider  a discrete spectrum, assume the special modes are light (\textit{i.e.} are  zero modes) and evaluate their low-energy 4D effective theory.

 In the presence of a bulk interaction, \textit{e.g.} a four-fermion operator with coefficient $\lambda$, with $[\lambda]=-3$, the coefficient of the effective 4D four-fermion operator 
 \be
  {\cal L}^{\rm eff}_{\rm 4D }=\lambda_{\rm 4}
  (\bar \psi \psi)(\bar \psi \psi) +\ldots\quad\quad \quad \, 
 \ee
 is given by
 \be
 \lambda_{\rm 4}= \lambda \int dy [f^{\Psi}_0(y)]^4\approx \lambda |m_\Psi|\,. \label{eq:lambdapsi4}
 \ee
 The coefficient grows with $m_\Psi$. This implies that if one tries to send $m_\Psi$ to infinity at fixed $\lambda$, the cutoff of the 4D EFT, which is roughly of order $\Lambda_{4} \sim 1/(4\pi \sqrt{\lambda m_\Psi})$, tends to zero. Hence taking such limit leads to a 4D EFT with vanishing  range of validity. 
 This is of course an obstruction to reach an exactly localized theory--- which  has finite range of validity. 
 However this argument still has a caveat because one could in principle adjust the magnitude of the 5D interactions (such as $\lambda$ in Eq.~\eqref{eq:lambdapsi4}) to keep the low-energy couplings under control. Unwanted interactions---for instance brane-localized ones---can be set to zero at a given scale if needed.

 To see where an obstruction  occurs with no possible caveat,   let us keep  evaluating the low-energy EFT of zero modes. In  the low-energy EFT the Kaluza-Klein modes are integrated out and contribute  as higher-dimensional operators in the low-energy EFT. In particular, in the presence of gravity, KK gravitons are  integrated out. The KK gravitons couple to the zero mode stress energy tensor, which contains terms proportional to bulk and brane masses.\,\footnote{ Here the bulk and brane masses are tied to each other to maintain a small 4D mass for the zero mode. } An example of scalar effective operator generated by the gravitons is ${\cal O}_\phi=(\partial_\mu \phi)^2 \phi^2$.

When the bulk and brane masses are sent to infinity,  the coefficient of this operator must tend to infinity unless cancellations occur to render the coefficient finite. An explicit calculation in AdS from \cite{Dudas:2012mv} shows that the coefficient of  ${\cal O}_\phi$ does diverge. Quoting their result, 
\be
{\cal L}_4= -\frac{e^{2 k L}}{6M^2_{\rm Pl}}\frac{(1+\alpha)^2}{3+2\alpha} {\cal O}_\phi \approx
-\frac{m_\Phi e^{2 k L}}{12 k M^2_{\rm Pl}} {\cal O}_\phi
\, \label{eq:L4_grav}
\ee
where we have taken the large $m_\Phi$ limit in the last step. Since for $m_\Phi \rightarrow \infty$ the coefficient of the effective operator tends to infinity,  
 the cutoff of the 4D EFT goes to zero. Hence the validity  range of the 4D EFT vanishes and the large bulk mass limit cannot continuously lead to an exactly localized  EFT. No coupling can be tuned here, since the strength of the interaction only depends  on $M_{\rm Pl}$. 
 
  Importantly, the obstruction occurs because of the presence of the KK gravitons.   The KK modes of fields other than gravity produce 4D effective operators which remain finite in the $m_\Phi \rightarrow \infty$ limit. 
We can now see what is special about gravity: since 5D gravity couples to 5D mass, taking the  $m_\Phi \rightarrow \infty$ limit would imply infinitely strong coupling  and thus no weakly coupled EFT description at any scale.

\subsection{Brane kinetic localization} \label{se:BKT}

A localized 4D mode for any of the matter fields $(\Phi, \Psi, {\cal A}^M)$ can also appear by taking the limit of a large brane kinetic term with  magnitude $r$ (see Sec.~\ref{se:BKTdir}). 
Aspects of the localization limit for each kind of field will be discussed further below. 
A general argument showing that the exact localization limit of  $\Gamma$ does not lead  to  $\tilde \Gamma$ is as follows.

At large $r$,  while one special  mode tends to get exactly localized on the brane, the set of all KK modes is fully expelled from the brane. The KK modes decouple from the brane, but certainly not from the spectrum. Rather, at large $r$ the set of KK modes gets a Dirichlet boundary condition on the brane.
This general feature can be seen explicitly in the dressed propagator Eq.~\eqref{eq:dressing2}, which  in the presence of a brane-localized kinetic term takes the form 
\be
   \Delta_p(y,y')= i\frac{\hat K_p(y) \hat K_p(y')}{\hat \Pi(p)+r p^2} +  \hat \Delta^D_p(y,y') \,. \label{eq:dressing2BKT}
  \ee
For $r\rightarrow\infty$, the bulk contribution $\hat \Pi(p)$ becomes negligible compared to $ r p^2$. The first term in Eq.~\eqref{eq:dressing2BKT} takes the form of a pure 4D pole. The second term corresponds to the set of Dirichlet KK modes, which clearly remain in the spectrum since they are not affected by the brane dressing. 
We consider a scalar propagator here, but the property remains valid for any kind of field. This feature  matches the well-known results from \cite{Carena:2002dz} about ``opaque'' branes.

\begin{figure}[t]
	\centering
	\includegraphics[width=0.2\linewidth,trim={0cm 0cm 0cm 0cm},clip]{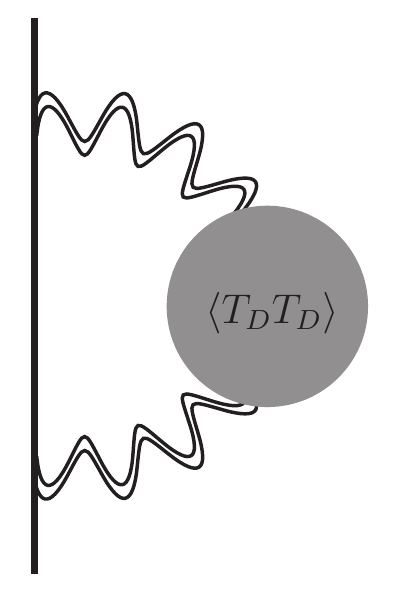}
	\caption{ Brane-localized modes interact with matter Dirichlet modes in the presence of gravity, therefore $\Gamma \neq \tilde \Gamma$.  $T_D$  denotes the  stress tensor of Dirichlet  modes. 
		}
	\label{fig:brane_grav_D}
\end{figure}

In the limit of exact localization, the classical part of the holographic action contains only a 4D field---with possible brane interactions  as discussed further below. 
Hence it may seem that the exact localization limit is indeed successful. 
However, at the quantum level, the brane localized  degree of freedom always know about the Dirichlet KK modes because of gravity. Namely, the brane modes couple to KK gravitons, which themselves couple to the Dirichlet KK modes of matter, as shown in Fig.~\ref{fig:brane_grav_D}. 

Hence the picture is that, while  $\tilde \Gamma$ contains by definition isolated 4D degrees of freedom, the same 4D degrees of freedom in  $\Gamma$ are necessarily accompanied by  towers of Dirichlet modes. Without gravity, the equivalence between  $\tilde \Gamma$ and  $\Gamma$ could be exact because the Dirichlet modes may be completely decoupled from the brane. In contrast, in the presence of  gravity, KK gravitons always connect brane modes to Dirichlet modes. This  has physically observable consequences therefore  the limit of $\Gamma$ at large $r$   differs from $\tilde \Gamma$.

Even though the above argument is sufficient to establish the discontinuity between the $\Gamma$ and $\tilde \Gamma$ theories in the presence of gravity, it is still interesting to study in more details the effects of kinetic localization for the various matter fields. This is relevant from a purely theoretical viewpoint but also for future model-building manipulations.  

\subsubsection{Scalar and fermion modes}

Since the $r$ coefficient normalizes the kinetic term, taking large $r$  has consequences for other operators of the  theory. At large $r$, canonical normalization implies the rescaling $\hat\phi=\sqrt{r}\phi$ and similarly for other fields,  which reduces by powers of $\sqrt{r}$ other operators of the brane Lagrangian. 
For scalar and fermions, it is always possible to obtain a massive, interacting Lagrangian in the $r\rightarrow \infty $ limit of $\Gamma$ by introducing brane-localized mass and interactions which scale with appropriate powers of $\sqrt{r}$. This does not change the fact that $\Gamma$ differs from $\tilde \Gamma$ by the presence of bulk Dirichlet modes.

\subsubsection{Gauge modes}

Kinetic localization of gauge fields is more constrained because, unlike in the case of scalar and fermions,  self-interactions of the gauge field are constrained by gauge invariance. As a result, when the coefficient of the gauge kinetic term grows, gauge self-interactions are necessarily suppressed. 

The gauge action reads 
\be
S_A = \int d^5X \sqrt{g}\left(
-\frac{1}{4g^2_5} {\cal F}^{MN}{\cal F}_{MN} \right)  + \int d^4x  \sqrt{|\bar g|}\left(
-\frac{r}{4 g^2_5} {\cal F}^{MN}{\cal F}_{MN} \right)\bigg|_{y=y_0}+ \ldots \label{eq:5Dgauge}
\ee
Consider  the transverse part of the propagator for ${\cal A}^\mu$, 
\be
\Delta^{\cal A}_{\mu\nu}(p;y,y')=\left(\eta_{\mu\nu}-\frac{p_\mu p_\nu}{p^2}\right) \Delta^{\cal A}(p;y,y') +\ldots
\ee
where in the presence of the brane kinetic term Eq.~\eqref{eq:5Dgauge}, 
\be
   \Delta^{\cal A}_p(y,y')= -ig_5^2\frac{\hat K_p(y) \hat K_p(y')}{\hat \Pi(p) + r p^2} +  \hat \Delta^{{\cal A},D}_p(y,y') \,. \label{eq:gaugerel}
  \ee

We can see that at large $r$ the self-energy takes the form 
\be
\frac{1}{g^2_5}\Pi(p)=\frac{1}{g^2_5}\left(\hat\Pi(p)+r p^2\right) \rightarrow \frac{r}{g^2_5} p^2 \,.
\ee
Therefore the effective gauge coupling in the quasilocalized limit is\,\footnote{This formula includes the case of a gauge zero mode in a compact extradimension, for which $ g^2_4=\frac{g_5^2}{L+r}$ for any $r$. }
\be
g^2_4\approx \frac{g_5^2}{r} \,.
\ee
This matching relates brane localization to the strength of  gauge interactions, and has thus important consequences. 

In order to achieve an exactly localized gauge field for a given value of the effective gauge coupling $g_4$, the increase of $r$ has to be accompanied with an increase of 
the 5D gauge coupling $g_5$. However, increasing $g_5$ has a price. Since $g_5$ controls 5D interactions, increasing it \textit{lowers} the 5D cutoff of the theory.
This implies that taking $r\rightarrow \infty$ at finite $1/g^2_4$ sends the cutoff of the theory to zero. The theory has thus a vanishing validity  range and cannot continuously reproduce the exactly localized gauge theory from $\tilde \Gamma$. Interestingly, in this case, the obstruction is not related to gravity.

Conversely, taking large $r$ for fixed $g_5$ and no requirement  on $g_4$, it seems one could obtain an exactly localized gauge theory with vanishing $g_4$ gauge coupling.  However, in the presence of gravity, this limit is obstructed by the Weak Gravity Conjecture (WGC) \cite{ArkaniHamed:2006dz}. In this limit the EFT cutoff is lowered to $\Lambda \sim g_4 M_{\rm Pl}$ as required by the WGC and thus taking $r\rightarrow \infty$ gives once again an EFT with vanishing validity  range. 
Hence there is again obstruction, in this case because of gravity.

\section{Braneworlds and Swampland \label{se:swamp}}

In the previous section we have shown that, at least in the presence of gravity, the exactly localized and quasilocalized braneworlds are not continuously related in theory space. 
In this section we focus on the exactly localized braneworld. We aim to find internal discrepancies or paradoxes in this kind of theory.

\subsection{Brane width \label{se:width}}

In the braneworld EFTs we consider, the brane is an infinitely thin  hypersurface. For an EFT without gravity ($M_{\rm Pl}\rightarrow \infty$), such feature can in principle remain valid at infinitely short distances.\,\footnote{If one removes gravity in the exactly localized theory, the bulk becomes totally empty and the fifth dimension can be trivially integrated over. }
In contrast, in the presence of gravity, the infinitely thin brane description should become invalid 
at distance scales of order of the local Planck length, where quantum fluctuations of spacetime become strong. 

From the EFT viewpoint, such  breakdown of the thin brane description should 
manifest itself via the presence of higher-dimensional operators encoding the effects of the brane width. 
These higher order terms in the braneworld EFT take the form
\be
S_{\rm brane}= \int d^5X \sqrt{|\bar g|} \left[\delta(y-y_0){\cal L}^{(0)}+
a\delta'(y-y_0){\cal L}^{(1)}+
\frac{b}{2}\delta''(y-y_0){\cal L}^{(2)}+\ldots
\right] 
\ee
where $a$, $b$ are coefficients vanishing in the $M_{\rm Pl}\rightarrow \infty$ limit.\,\footnote{
The brane  profile, taken as a distribution, can be formally expanded over the basis of the Dirac delta's derivatives. Truncation of this series  depends on the test function on which it acts. In the context of the low-energy EFT, this truncation  is controlled by the long-distance expansion defining the EFT. 
}
 Without any specification of the UV completion or of the exact brane profile, this immediately implies that the  ${\cal L}^{(i)}$ have to depend on $y$---otherwise all the  ${\cal L}^{(i>0)}$ would vanish. The fields in ${\cal L}^{(i)}$ are thus 5D fields, which implies that the theory is a quasilocalized braneworld---as defined in Eq.~\eqref{eq:Sgen}. 

In short, gravity requires that the brane has some concept of width, which requires all fields to be five-dimensional, such that the braneworld is of the quasilocalized kind. 
From the viewpoint of a UV completion this could for example happen because the brane is a soliton with finite width \cite{Rubakov:1983bb, ArkaniHamed:1999za, Mirabelli:1999ks},  or because  the brane becomes a dynamical object with a non-trivial form factor near the Planck scale. 

From this simple brane width argument we may conclude that an exactly localized braneworld EFT is incompatible with an embedding  into a theory of gravity. 
In the following we present further arguments,  relying in part on standard swampland conjectures (see \cite{Palti:2019pca} for a review).

\subsection{Argument from global symmetries \label{se:sym}}

Consider a flat 5D interval with a $U(1)$ gauge field in the bulk. Assume two species  $\phi_0$, $\phi_1$ with charges $q_0$, $q_1$  exactly localized on two different branes located at each  endpoints of the interval. To be specific we assume that $q_0$, $q_1$ are coprime and of opposite sign.

Let us  consider the low-energy theory below the KK scale, for which all KK photons and KK gravitons are integrated out. The low-energy limit is taken only for convenience, the argument still applies at any energy scale in the theory. 
The 4D effective Lagrangian contains effective operators generated by the KK modes. Because of exact localization, the 4D Lagrangian only contains operators
composed of mononimals $|\phi_0|^{2}$, $|\phi_1|^{2}$ 
and similar ones with derivatives and more complex Lorentz structures.

   In this 4D theory the $\phi_0$ and $\phi_1$ numbers $N_0$, $N_1$ are separately \textit{exactly} conserved. Conservation of these numbers is not implied by the gauge symmetry, which only dictates conservation of the gauge charge $q_0 N_0+q_1 N_1$, hence the individual $N_0$, $N_1$ numbers are global charges. 
  The theory has therefore an exact global symmetry.

This is in direct contradiction with the swampland conjecture that there is no exact global symmetry in an EFT emerging from a UV theory of  gravity. 
This contradiction is resolved in the quasilocalized picture, where $\phi_0$, $\phi_1$ are the zero modes of 5D bulk fields $\Phi_0$, $\Phi_1$. These bulk fields are directly in contact via 5D operators respecting the gauge symmetry but not the individual $\Phi$ number (see discussion in \cite{Fichet:2019ugl}). 
The zero modes of $\Phi_1$, $\Phi_2$, even if highly localized on each brane, have a non vanishing wavefunction in the bulk and thus overlap with each other. As a result, in addition to $|\phi_i|^2$ monomials, the low-energy theory  contains operators build from
monomials of 
\be
\phi_0^{q_1} \phi_1^{q_0}+{\rm h.c.}\,
\ee
which explicitly violate the individual $\phi_i$ numbers. These operators arises both from the direct contact between the zero modes and from integrating the KK modes of $\Phi_0$ and $\Phi_1$.  Such symmetry-violating terms would be absent in case of exact localization, causing the global symmetries to be exact.

Summarizing, we have  presented a configuration where  exact localization of charged fields is tied to a violation of the conjecture that no exact global symmetry exists in the presence of gravity. This violation is naturally avoided when using quasilocalized fields. 
 A similar argument has been recently presented in \cite{Fichet:2019ugl}.

\subsection{Argument from  emergent species \label{se:species}}

Consider a slice of AdS$_5$, \textit{i.e.} AdS$_5$ space truncated by two branes. 
This corresponds to $a(y)=ky$ in the metric Eq.~\eqref{eq:metric}. For AdS it is convenient to use the conformal coordinates $z=e^{ky}/k$. The branes are taken to be at positions $z_0=1/k$ (UV brane), $z_1=1/\mu$ (IR brane).   
For the moment we assume no matter on the IR brane or in the bulk. For an introduction to QFT in a warped background see \textit{e.g.} \cite{ Ponton:2012bi,Gherghetta:2010cj}).

While the cutoff in terms of proper distance is constant since AdS is homogeneous, the cutoff in coordinate distance varies along the $z$ coordinate. Assuming the 5D cutoff for an observer on the UV brane is $\Lambda$, the cutoff for an observer on the IR brane is $\Lambda'=\Lambda \mu/k$, \textit{i.e.} it is ``warped down''.\,\footnote{  This is because the effect of higher dimensional operators in the 5D action is enhanced by powers of $k/\mu$ on the IR brane as compared to the UV brane. }

Let us consider the holographic action defined on the UV brane---as usually done in the context of AdS/CFT. As well-known \cite{ArkaniHamed:2000ds, Polchinski:2002jw, Gherghetta:2003he, Fichet:2019hkg}, for 4-momentum $|p|=|\sqrt{p^2}| \gg \mu$, IR localized fields/operators and the IR brane itself vanishes from all correlators. In this 5D regime the theory can be effectively described by a UV brane and an infinite AdS bulk. 
Since the IR brane appears only in the IR \textit{i.e.} at low 4-momentum $|p|$, it is effectively emergent from the viewpoint of the UV brane, as formalised in the holographic action (see  \cite{Brax:2019koq,Costantino:2019ixl} for  BSM application). 
 To facilitate  discussions, let us just assume the transition is at $|p|\sim \Lambda'$ with $\Lambda'$ of order $\mu$. 
We consider the two extreme regimes of the holographic action. For $|p|\gg \Lambda'$, the theory is pure AdS$_5$. For $|p|\ll \Lambda'$, the theory  contains only zero modes and is 4D.

Let us then introduce exactly localized matter on the IR brane. We are free to add  a large number  of species $N\gg 1$, all exactly localized on the IR brane. 
Because of this large number of species, in the 4D regime, the cutoff is lowered to $\Lambda'/\sqrt{N}$ as dictated by the
species scale. The species scale is a swampland conjecture implied by gravity \cite{Dvali:2007hz,Palti:2019pca}. This introduces a rather strange feature.  In the holographic action, there is now a parametrically large energy range in between the 5D and 4D regimes, \be
|p|\sim [\Lambda'/\sqrt{N}, \Lambda' ] \label{eq:disc_Lambda}
\ee
 for which the EFT is invalid. We take this discontinuity as a signal of an inconsistency. 
 
 The feature is related to the emergence of many degrees of freedom in the IR. Such parametrically large increase of degrees of freedom is in gross disagreement with the picture that degrees of freedom should monotonically decrease when flowing towards the IR, as  encoded by c- and a-theorems. 
It would be interesting to evaluate explicitly the holographic $a(z)$ function along the lines of \cite{Freedman:1999gp,Myers:2010tj}. However for our purposes,  qualitative considerations are enough: The holographic action definitely has a problem with IR degrees of freedom.

Both inconsistencies about validity range and   IR degrees of freedom are solved when assuming quasilocalized fields instead of exactly localized fields. 
With quasilocalized fields, the theory now contains $N$ bulk fields. The holographic action knows about these bulk degrees of freedom at any $|p|$. The $N$ bulk fields imply an overall reduction of the 5D cutoff by $\sqrt{N}$ in both 5D and 4D regimes, and no discontinuity  in the validity range of the theory (in contrast with  Eq.~\eqref{eq:disc_Lambda}).   The existence of the $N$ bulk fields being  known in the UV, no steep increase in the number of degrees of freedom
due the emergent IR brane occurs along the RG flow.

Let us comment on the interplay with gravity. 
The cutoff-based argument  relies on the species scale, which is implied by gravity.
The argument about degrees of freedom seems naively unrelated to gravity, although this may deserve further thinking since the evaluation of the usual holographic $a$ function does rely on Einstein's equations.\,\footnote{
One can also argue that the presence of the $N$ 5D fields is implied by the finite IR brane width as discussed in Sec.~\ref{se:width}, and thus enforced by gravity.
}

Summarizing, in the warped configuration  studied here, exact localization of a large number of species leads to inconsistencies which are partly related to the presence of gravity. These inconsistencies are naturally solved when fields are taken to be quasilocalized.

\subsection{Discussion}

We have exhibited two specific models with exactly localized fields which, to the best of our understanding, should belong to the swampland. 
We have also made the simple point that whenever some notion of brane thickness is introduced, the braneworld should be of the quasilocalized type. 
These points are unfavorable to the  exactly localized braneworld EFT. 

On the string theory side, braneworld model-building is often done with D3-branes, which give rise to   matter fields living strictly on the  worldvolume (see \textit{e.g.} \cite{Aldazabal:2000sa,Antoniadis:2000jv,Uranga:933469}). This may seem to favor, at first view, the  picture of an exactly localized braneworld---which stands in contrast to the observations made in the rest of this section. 
However a full string picture has restrictions, for instance  D3-branes have to be accompanied by D7-branes wrapped around compact space dimensions. The D7-branes do generate a tower of matter KK modes, which somehow accompany the isolated states from D3-branes. 
The presence of  matter KK modes  would then be reminiscent of the quasilocalized picture. 
Also, a notion of thickness for the D-brane  is sometimes discussed in the literature \cite{Moeller:2000jy}. This would again imply that the low-energy limit has to be a quasilocalized braneworld.

Given the possible subtleties on the string side, we do not attempt a broad conjecture about the  (field-theoretical) exactly localized braneworld.  The precise string picture relative to exact/quasi-localization would deserve a detailed study. 
Here we simply report our results with no further extrapolation.

All these considerations about exactly versus quasi-localized braneworld are interesting from a conceptual viewpoint, but also have concrete observable consequences as we will see in next section.

\section{ The quasilocalized warped braneworld } \label{se:RS}

Given the previous results, it is interesting to revisit  existing braneworld models  of the exactly localized kind. 
This includes in particular the DGP braneworld \cite{Dvali:2000hr} and the  Randall/Sundrum II (RSII) braneworld \cite{Randall:1999vf}, both originally presented with the SM exactly localized  on a brane. 

In a sense, an exactly localized braneworld is an approximation of a quasilocalized one. How good  is the approximation may depend on the spacetime background, on the field content and so on. 
As a general tendency, we can expect a richer phenomenology once matter is quasilocalized,  since new degrees of freedom (the KK modes) are always present in the theory, and since a quasilocalized brane field has direct contact with bulk degrees of freedom. 
Taking into account these phenomena may provide new observable effects, and perhaps new constraints on the braneworld model.

In this work we focus on the ``quasilocalized RSII model'', \textit{i.e.} RSII where all SM fields are quasilocalized. 
We include an IR brane to discretize the spectrum, as it is sometimes convenient for discussions. The IR brane can be sent to infinity at any time to recover  full AdS space in the IR. 

For every localized 4D field, there is a KK tower, or a KK continuum if the IR brane is at infinity. 
The phenomenology for scalar and fermions depends both on their bulk mass and their brane localized Lagrangian---which are responsible of the two localization mechanisms discussed in Secs.~\ref{se:limit}, \ref{se:disc}. In contrast,  quasilocalized gauge fields are much more constrained because 5D gauge symmetry constrains their profile and their interactions. 
The phenomenology (including possible constraints) from the scalar and fermion KK sectors is certainly interesting, but our focus here is on the gauge and gravity sectors which are more model-independent.

\subsection{Action, propagator, opacity and EFT validity}

Consider the 5D action of  gravity and a gauge field. The action takes the form 
\be 
S_{\rm AdS}= \int d^5X \sqrt{g}\left[ M^3_* {\cal R} - \Lambda_5
-\frac{1}{4g^2_5} {\cal F}^{MN}{\cal F}_{MN}
\right]
  + \int_{\rm br.} d^4x  \sqrt{|\tilde g|}\left(
-\frac{r}{4 g^2_5} {\cal F}^{MN}{\cal F}_{MN}-\Lambda_4 \right)
\,.
\label{eq:RS_action}
\ee
The 5D  cosmological constant and brane tension satisfy $\Lambda_5=-12 k^2 M^3_*$, $\Lambda_4=\Lambda_5/k$, $k$ being the AdS curvature. The $M_*$ parameter sets the strength of 5D gravity
 and is related to the 4D Planck mass by $M_*^3\approx kM^2_{\rm Pl}$. 
The metric of the AdS background is denoted $\gamma_{MN}$, such that $g_{MN}=\gamma_{MN}+\ldots$ where the ellipse denotes the metric fluctuations. The graviton Lagrangian will be expanded in Sec.~\ref{se:AAAA}. 

A localized Ricci scalar could also be included on the brane. Since our focus is on matter fields, this is  a  direction  we do not consider in the scope of this work. 
Optionally, another brane with tension $-\Lambda_4$ and no localized matter Lagrangian  is also  included in the action Eq.~\eqref{eq:RS_action}, further away from the AdS boundary, \textit{i.e.} in the IR region. This second brane is referred to as  ``IR brane'' and the main one ``UV brane''.

For AdS$_5$ the general metric of Eq.~\eqref{eq:metric} satisfies $a(y)=ky$. We switch to so-called conformal coordinates $z=e^{ky}/k$, giving
\be
ds^2=\gamma_{MN}dX^MdX^N=(kz)^{-2}(\eta_{\mu\nu}x^\mu x^\nu-dz^2)\, \label{eq:metricz}
\ee
where $\eta_{\mu\nu}$ is Minkowski metric with $(+,-,-,-)$ signature. The UV brane is taken to be at $z=z_0=1/k$ with no loss of generality.  The IR brane  is situated at $z=z_1=1/\mu$. 

To disentangle the components of the 5D gauge field, one introduces the 5D gauge fixing functional
\be
-\frac{1}{2\xi k z g_5^2}\left(
\partial^\mu {\cal A}_\mu-\xi z\partial_5 \left(z^{-1}{\cal A}_5 \right)
\right)^2\,,
\ee
defining the $R_\xi$ gauge \cite{Randall:2001gb,Carena:2002dz}. For our purposes it is enough to work in the Feynman gauge $\xi=1$. 
The $\langle{\cal A}_5{\cal A}_5\rangle $ propagator encodes the longitudinal degrees of freedom. 
The ${\cal A}_\mu$ component of the gauge field is taken to satisfy Neumann boundary condition on the branes while  ${\cal A}_5$ has Dirichlet boundary conditions. 
 The propagator for ${\cal A}_\mu$ in the presence of the IR brane  reads
 \begin{align}
&\langle{\cal A}_\mu(p,z){\cal A}_\nu(-p,z')\rangle  =
\Delta^{\cal A}_p(z,z')=
\\ \nonumber
 & - \eta_{\mu\nu}\,
i\frac{\pi k^3 (zz')^2}{2 }   
\frac{
\left[ Y_{0}\left(p/k\right)J_{1}\left(pz_<\right)
-  J_{0}\left(p/k\right) Y_{1}\left(pz_<
\right)\right]\left[
 Y_{0}\left(p/\mu\right)J_{1}\left(pz_>\right)
-  J_{0}\left(p/\mu\right) Y_{1}\left(pz_>
\right)
\right]}
{ J_{0}\left(p/k\right)   Y_{0}\left(p/\mu\right)
-   Y_{0}\left(p/k\right)  J_{0}\left(p/\mu\right)}\,
\end{align}
where $p=\sqrt{\eta^{\mu\nu} p_\mu p_\nu}$.  

The 5D action Eq.~\eqref{eq:RS_action} is the leading term of a low-energy  effective theory  valid at distances larger than $\Delta X \sim 1/\Lambda$ where $\Lambda$ is the  validity cutoff. 
The cutoff is set by the strongest interaction, \textit{i.e.} either by gravity  or by gauge interactions, giving respectively
\be
M^3_*\sim \frac{\Lambda^3}{24 \pi^3}\,,\quad\quad \frac{1}{g^2_5}\sim \frac{c\Lambda}{24 \pi^3} \,\label{eq:NDA}
\ee
where $c$ is a group theoretical factor of order of the number of colors \cite{Chacko:1999hg,Agashe:2007zd}. The gravity cutoff implies $k \lesssim M_{\rm Pl}$ for the higher order curvature terms to be negligible. 
The coupling of KK gravitons is controlled by the dimensionless quantity
\be
\kappa=\frac{k}{M_{\rm Pl}} \,
\ee
which can go up to $O(1)$. 

In the coordinates Eq.~\eqref{eq:metricz}, the cutoff on $p$ as seen by a local observer at position $z$ is $\Lambda\,kz$. Hence for a given momentum $p$, the EFT breaks down when going far enough in the IR region, at roughly $z =O( 1/p) $ (see \textit{e.g.} \cite{Goldberger:2002cz}). 
However a property of the propagators is that they tend to be exponentially suppressed when an endpoint enters this IR region \cite{ArkaniHamed:2000ds, Polchinski:2002jw, Gherghetta:2003he, Fichet:2019hkg}. This is true for both Euclidian and Lorentzian momentum. For Lorentzian momentum the suppression appears once the propagator is dressed by bulk interactions. One has
\be
\Delta_p(z) \sim \begin{cases}  
e^{- |p| z_>} \quad &{\rm if}~~p_\mu~\quad{\rm spacelike}
\\
 e^{-C p z_>} \quad &{\rm if}~~p_\mu~\quad{\rm timelike}
 \end{cases} \label{eq:Kexp}
 \ee
 An analytical estimate near strong coupling gives typically $C\sim O(1)-O(0.1)$.
The holographic profiles are expressed in terms of propagators (Eq.~\eqref{eq:K_class}) hence the same property is true for them. 
 This opacity property of AdS tends to censor  the IR region where the 5D EFT breakdowns, \textit{e.g.} where gravity would become strongly coupled. We will see an example of calculation relying on the cutoff from Eq.~\eqref{eq:Kexp} in Sec.~\ref{se:AAAA}.

\subsection{Anomalous running of gauge couplings}
\label{se:RG}

\begin{figure}[t]
	\centering
	\includegraphics[width=0.5\linewidth,trim={0cm 0cm 0cm 0cm},clip]{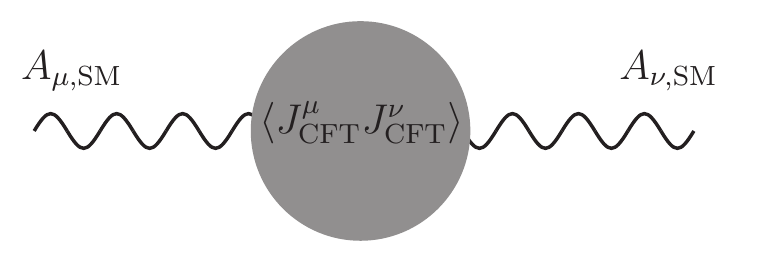}
	\caption{ SM gauge fields dressed by insertion of CFT correlators, equivalent to the effect of the gauge KK continuum on brane-localized SM fields. 
		}
	\label{fig:gauge_CFT}
\end{figure}

We now treat the gauge field holographically, introducing the variable
\be
{\cal A}_{\mu,0}={\cal A}_\mu\big|_{z=z_0}\,.
\ee
Using asymptotic forms of Bessel functions, the bilinear holographic action is found to be
\begin{align}
 \label{eq:RGlargep}
\Gamma_{\rm cl}[ {\cal A}_{\mu,0}]& \approx 
\begin{dcases}
\int \frac{d^4p}{(2\pi)^4}\left(\frac{\log\left(k/\mu\right)}{k}+ r \right)  \frac{p^2}{4g^2_5} {\cal A}_{\mu,0}(p){\cal A}^\mu_{0}(-p)  \\
  \int \frac{d^4p}{(2\pi)^4}\left(\frac{\log\left(2k/\sqrt{-p^2}\right)-\gamma}{k}+ r \right)  \frac{p^2}{4g^2_5} {\cal A}_{\mu,0}(p){\cal A}^\mu_{0}(-p) 
\end{dcases}
\end{align}
For $|p|<\mu$, the  action  matches the one of gauge zero modes, and the low-energy gauge coupling $g_4$ takes the constant value
\be
\frac{1}{g^2_{4,0}}= \frac{1}{g_5^2} \left(\frac{\log\left(k/\mu\right)}{k}+ r \right) \,. \label{eq:g4LE}
\ee
For $|p|>\mu$, we can see that the holographic action is non-analytic. This regime includes the case of no IR brane $\mu \rightarrow 0$. In this regime the action describes a running holographic gauge coupling
\be
\frac{1}{g^2_4(p)}= \frac{1}{g_5^2} \left(\frac{\log\left(2k/\sqrt{-p^2}\right)-\gamma}{k}+ r \right) \,.
\label{eq:g4HE}
\ee
Combining Eqs.~\eqref{eq:g4LE},~\eqref{eq:g4HE}, neglecting the small term $\log(2)-\gamma$ for simplicity,  
and  using $r k \gg \log(k/\mu)$ which is the regime of relevance for our discussion, we get
\be
g^2_4(p)=\frac{g^2_{4,0}}{1-\frac{1}{2rk}\log(-p^2/\mu^2)} \label{eq:g4RG} \,.
\ee
Here we have expressed the running in term of the low-energy coupling $g_{4,0}$, but we could similarly define the running at any scale $p_0$ and obtain a similar form.

We obtain the well-known feature that the AdS bulk dynamics induces a tree-level running of the holographic gauge coupling \cite{Contino:2002kc,Randall:2001gb,Goldberger:2002hb}. This running is induced by the presence of the KK continuum. Because of AdS/CFT, the running is equivalently described by mixing a 4D gauge field to a conserved current of the CFT. This produces exactly the same effect, and can be understood as dressing the gauge field by loops of the CFT constituents---which indeed contribute to the beta function of $g_4$. The fact that a tree-level effect on the AdS side matches a loop effect on the CFT side is also  understood \cite{Aharony:1999ti}.

Let us now consider this behaviour in the context of a quasilocalized warped braneworld, where gauge fields  as shown above are identified with SM gauge fields, $A^{\rm SM}_\mu\equiv{\cal A}_{\mu,0}$. 
In that context the presence of the bulk dynamics (the gauge KK continuum) induces an anomalous tree-level running of the SM gauge couplings. Using AdS/CFT, this effect can equivalently be understood as the mixing to a current from a hidden conformal sector. 

Clearly, such anomalous running has to be small otherwise it would have already been observed. From the running shown in Eq.~\eqref{eq:g4RG}, we can see that
the condition for the effect to be small over a range of energy $[p_0^2,p_1^2]$ is 
\be
\log(p_1^2/p_0^2)\ll rk\,. \label{eq:g4bound}
\ee

How can this be realized in the model? Let us focus on the $r$ parameter. 
Because of gauge symmetry, the gauge sector is very constrained and $r$ is the only free parameter. Moreover, for a given value $\bar g_4$, \textit{e.g.} $\sim 1/137$, or more generally the typically value of $g_4$ over $[p_0^2,p_1^2]$, the brane contribution $r/g_5^2$ is bounded from above, as can be seen from Eq.~\eqref{eq:g4LE} or \eqref{eq:g4HE}. This brane contribution can be at most as large as $1/\bar g^2_4$,
\,\footnote{We do not consider the fine-tuned case of a negative $r$  cancelling  the bulk contribution to high precision.   }
\be
\frac{r}{g_5^2} < \frac{1}{\bar g^2_4}\,. 
\ee
It follows that the only way to increase $r$  is to simultaneously \textit{increase} $g_5$. 

The other way to satisfy Eq.~\eqref{eq:g4bound}  would be to increase $k$. However $k$ is bounded from above since $k\lesssim M_{\rm Pl}$.  $k$ controls the strength of graviton coupling.  Hence  we obtain again   that a bound on the anomalous running of the gauge coupling will constrain the \textit{weak} values of a coupling---which is the opposite of how usual experimental bounds work. 
This implies that the parameter space of the braneworld  can be cornered such that the model could---in principle---be tested completely. 

To see this, let us return to the $g_5$ coupling. If the cutoff of the 5D theory is set by $g_5$ (see Eq.~\eqref{eq:NDA}), requiring larger $g_5$ implies a lower EFT cutoff $\Lambda$.  In terms of $g_5$ this is given by Eq.~\eqref{eq:NDA} and in terms of $r$ this is given by  $1/r\sim g_4^2 c\Lambda / (24\pi^3)$. 
On the other hand, conventional high-energy experiments should bound the cutoff $\Lambda$ from below, which is just the usual experimental situation. Therefore $\Lambda$ can in principle bounded from both above and below. 

Summarizing, avoiding a large anomalous running of gauge couplings in the warped quasilocalized braneworld amounts to require stronger
coupling of bulk degrees of freedom.\,\footnote{In the qualitative ``compositeness'' language, this amounts to say that the mixing between the elementary fields and the composite sector is suppressed when the composite sector has stronger self-interactions $g_5$. }
This effect is specific to the gauge sector, where gauge symmetry ties together localization  and strength of interactions.

\subsection{Anomalous gauge boson scattering from 5D gravity}
\label{se:AAAA}

\begin{figure}[t]
	\centering
	\includegraphics[width=0.2\linewidth,trim={0cm 0cm 0cm 0cm},clip]{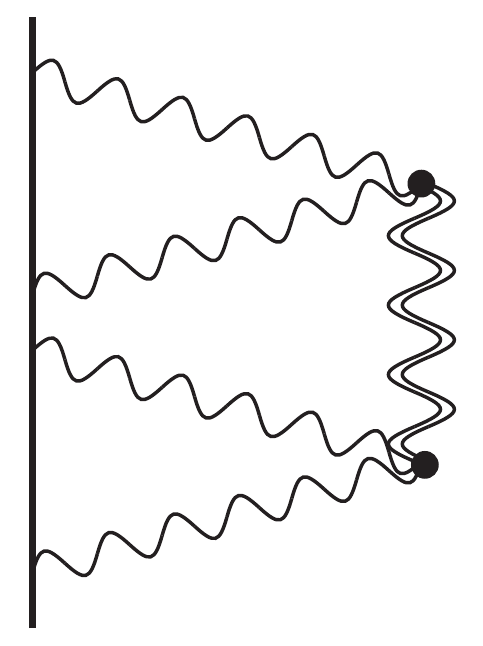}
	\caption{ 
	Gauge boson scattering induced by 5D gravitons. 	}
	\label{fig:brane_gauge_grav}
\end{figure}

 In the quasilocalized braneworld, the gauge bosons have a fraction of their wavefunction living in the bulk.
 Unlike  the exactly localized case , 
 the gauge fields can thus be directly in contact with \textit{e.g.} 5D gravity. 

The relevant interaction is encoded in the kinetic term
  \be
-\int d^4x dz \sqrt{-g} \, \frac{1+ r \delta(z-z_0)}{4 g_5^2} {\cal F}_{MN} {\cal F}^{MN} \,. \label{eq:gaugekinRS}
\ee
The distribution of the gauge fields between bulk and brane can be read off this kinetic term---when setting the metric $g_{MN}$ to the background value $\gamma_{MN}$. 
 We can notice that the bulk component would tend to zero for $r\rightarrow \infty$. However, in the case of gauge bosons, large $r$  requires to take large $g_5$, which is constrained as discussed in Sec.~\ref{se:RG}.
 
 The coupling of the 5D graviton to the gauge field can be derived from Eq.~\eqref{eq:gaugekinRS} by expanding the metric as \be g_{MN}=\gamma_{MN}+
 \sqrt{\frac{2}{M^3_*}}h_{MN}+\ldots
 \ee
Expanding the Ricci scalar at quadratic order gives the graviton kinetic term ${\cal L}_h$ and the relevant action reads
 \be
S_h = \int d^4x dz \sqrt{-\gamma} \left(
 {\cal L}_h + \sqrt{\frac{1}{2M^3_*}}h^{MN} T_{MN}\right) \,.
 \ee
 The full graviton kinetic term can be found in \textit{e.g.} \cite{Boos:2002hf,Hinterbichler:2011tt, Dudas:2012mv}. 
   The stress tensor for the gauge field reads
\be
T_{MN} =  
\frac{1+ r \delta(z-z_0)\delta_{M\mu}\delta_{N\nu}}{ g_5^2}  \left(- {\cal F}_{MV} {\cal F}_{N}^{\,\,V} +\frac{1}{4}\gamma_{MN}{\cal F}_{PQ}{\cal F}^{PQ}
\right)
\,.
\ee

The 5D gravitons induce a tree-level scattering of the  gauge bosons. In our holographic formalism this is encoded in the holographic 4-point function $\langle
{\cal A}_{\mu,0}{\cal A}_{\nu,0}{\cal A}_{\rho,0}{\cal A}_{\sigma,0}
\rangle$.
Our interest here is in the big picture, we want to obtain the parameter dependence of the amplitude. 
We will not give the detailed structure of the
graviton-induced gauge boson scattering. These can be found in \textit{e.g.} \cite{Fichet:2014uka}. Also we focus only on the contribution from the spin-2 helicity degrees of freedom. 

Following \cite{Dudas:2012mv},  the graviton degrees of freedom can be disentangled using field redefinitions and appropriate gauge fixing. 
 The diagonal helicity-2 degrees of freedom are given by the traceless part of $(kz)^2 h_{MN}$, noted $ \tilde h_{\mu\nu}$,\,\footnote{Namely 
 $ \tilde h_{\mu\nu}=\hat h_{\mu\nu}-\frac{1}{4}\eta_{\mu\nu} \hat h^\rho_\rho$, $\hat h_{MN}=(kz)^2 h_{MN}$.   }
which couples to the source
\be
\tilde T_{\mu\nu}=T_{\mu\nu}-\frac{1}{4}\eta_{\mu\nu}T_\rho^\rho\,. 
\ee
The relevant piece of the graviton action is
\begin{align}
S^h=\int d^4xdz\left( 
\frac{1}{2 (kz)^3}(\partial_R \tilde h_{\mu\nu})^2
 +\frac{1}{\sqrt{2 M_*^3}}
\frac{1}{(kz)^3} \tilde h^{\mu\nu}\tilde T_{\mu\nu}
  \right)\,. \label{eq:Lagh}
\end{align}
In Eq.~\eqref{eq:Lagh} all contractions are done with the Minkowski metric. The $\tilde h_{\mu\nu}$ component has Neumann boundary conditions on the branes. 

The exact graviton propagator is $\langle\tilde h_{\mu\nu} \,\tilde h_{\mu'\nu'}\rangle = \eta_{\mu\nu}\eta_{\mu'\nu'} \Delta^{\bm 2}_p(z,z')$  with
\begin{align}
& \Delta^{\bm 2}_p(z,z')=  \\ \nonumber
 & i\frac{\pi k^3 (zz')^2}{2 }   
\frac{
\left[ Y_{1}\left(p/k\right)J_{2}\left(pz_<\right)
-  J_{1}\left(p/k\right) Y_{2}\left(pz_<
\right)\right]\left[
 Y_{1}\left(p/\mu\right)J_{2}\left(pz_>\right)
-  J_{1}\left(p/\mu\right) Y_{2}\left(pz_>
\right)
\right]}
{ J_{1}\left(p/k\right)   Y_{1}\left(p/\mu\right)
-   Y_{1}\left(p/k\right)  J_{1}\left(p/\mu\right)}\,.
\end{align}
The propagator is exponentially suppressed in the IR region, as described in Eq.~\eqref{eq:Kexp}. In the  $z_><1/|p|$ region, it takes the form
\be
\Delta^{\bm 2}_p(z,z') \approx
 i\frac{2k}{p^2}+i\frac{2\gamma-1+2\log\left(\sqrt{-p^2}/2k\right)  }{2k}
-i\frac{\left((kz_<)^2-1\right)^2}{4k} \label{eq:Deltagrav2} \,.
\ee
This is the region of interest. 
Here we have taken the continuum limit such that the poles do not appear.\,\footnote{As shown in e.g. \cite{Fichet:2019hkg}, the KK  modes get a width from dressing by bulk interactions, tend to overlap with each other and give rise to a branch cut---corresponding to the AdS continuum.  }  The zero mode in Eq.~\eqref{eq:Deltagrav2} corresponds to the 4D graviton. The second term encodes the effect of the KK continuum on the UV brane \textit{e.g.} the correction to the Newton potential. 
The last term is the Dirichlet contribution, as shown in the form  Eq.~\eqref{eq:dressing2}. This Dirichlet term is the leading one in the physical process we consider.

Let us now consider the scattering of four on-shell  gauge boson. For on-shell massless gauge bosons the holographic profiles are simply $1$ for any $z$. 
The scattering is induced at tree-level by  graviton exchange. 
Using that $K=1$, the relevant stress tensor expressed with the holographic variables is
\be
\tilde T_{\mu\nu}=\frac{1+ r \delta(z-z_0)}{ g_5^2}  (kz)^2 \left(- {\cal F}_{\mu \rho,0} {\cal F}_{\nu,0}^{\,\,\rho} +\frac{1}{4}\eta_{\mu\nu}{\cal F}_{\rho \sigma,0}{\cal F}_0^{\rho \sigma}
\right)
\ee
where contractions are done with the Minkowski metric. 

The polarization structure is encoded in the tensor
\begin{align}
&{\cal E}^{\mu\nu}(12)= \\
&\frac{1}{2}\left(p_1^\mu p_2^\nu \, \epsilon_1 . \epsilon_2 + 
\epsilon_1^\mu \epsilon_2^\nu \, p_1 . p_2- 
p_1^\mu \epsilon_2^\nu \, \epsilon_1 . p_2-
p_1. \epsilon_2 \, \epsilon^\mu_1  p^\nu_2
+1\leftrightarrow2\right)
-\eta^{\mu\nu}\frac{1}{2} (p_1. p_2 \, \epsilon_1 . \epsilon_2 - p_1.\epsilon_2\, p_2.\epsilon_1  )\,
\end{align}
here defined for two ingoing gauge bosons with momentum $p_1$, $p_2$ and polarization vectors $\epsilon^\mu_1$, $\epsilon^\mu_2$.  Properties of the helicity amplitudes from spin-2 exchange can be found in \textit{e.g.} \cite{Fichet:2014uka} and need not be discussed here.

To get a familiar form for the amplitude we have to use canonically normalized  external states. Starting from the holographic fields ${\cal A}_0$, this is done by including a factor $g_4(Q)$ for each external gauge boson leg. The $g_4(Q)$ is defined in Eq.~\eqref{eq:g4RG}. Here $Q$ is some typical scale involved in the physical process. Since we are interested in the large $r$ limit, this tree-level running effect is irrelevant and we simply take $g_4\approx g_5/\sqrt{r}$.

Putting everything together, the amplitude takes the form
\be
i{\cal M}(12\rightarrow 34)=i{\cal M}^s+i{\cal M}^t+i{\cal M}^u\,.
\ee
with
\be
i{\cal M}^s=\frac{2}{M^3_*}{{\cal E}}_{\mu\nu}(12){{\cal E}}^{\mu\nu}(34)
\int dz dz' \frac{1}{kz kz'}\frac{1+ r \delta(z-z_0)}{ r}\frac{1+ r \delta(z'-z_0)}{ r}
\Delta^{\bm 2}_s(z,z')
\ee
and similarly for the $t$ and $u$ diagrams.

Let us consider  the pure AdS regime $\sqrt{s}>\mu$. 
The propagator is exponentially suppressed in the IR region as dictated by opacity in the timelike region, see Eq.~\eqref{eq:Kexp}. For simplicity we do not take into account the $C$ coefficient from the exponential, and assume suppression in the  
$z_>\sim 1/\sqrt{s}$ region. The same region can be taken for the position integral of the cross diagrams. 
The non-vanishing contribution to the position integrals comes from the $\sqrt{s} < 1/z_>$ region of momentum space where the propagator takes the form 
Eq.~\eqref{eq:Deltagrav2}. 
The leading contribution is found to be
\be
i{\cal M}^s \approx   \frac{\kappa^2}{8\, k r\,s^2} {\cal E}_{\mu\nu}(12){\cal E}^{\mu\nu}(34) + \ldots
\label{eq:AAAA}
\ee

This main contribution comes from the Dirichlet piece of the graviton propagator. The ellipse represents subleading  contributions. 
The amplitude is of course controlled by the 5D gravity strength $\kappa$. Interestingly, it turns out that this amplitude  is scale invariant.

We can see that the amplitude is suppressed  by $r$, \textit{i.e.} the more the gauge fields are brane localized the less they see 5D gravity. 
 For a given coupling $g_4$, large $r$ can only be accomplished with large $g_5$. 
Hence as in the previous subsection, we see that an upper bound on this new physics effect amounts to lower the new physics cutoff. 

Since the scattering amplitude Eq.~\eqref{eq:AAAA} is scale invariant, it can be tested on an equal footing by experiments at very different scales. This scale invariance should certainly have interesting consequences regarding the interplay between different experiments. 

Finally, if an IR brane exists and $\sqrt{s} < 1/\mu$, all KK modes are effectively heavy and give rise to a local amplitude
\be
i{\cal M}^s \approx   \frac{\kappa^2}{16 (kr)^2 \mu^4 } {\cal E}_{\mu\nu}(12){\cal E}^{\mu\nu}(34) + \ldots
\label{eq:AAAALE}
\ee
This amplitude can also  be described by a 4D EFT with two local Euler-Heisenberg operators (see \textit{e.g.} \cite{Fichet:2014uka,Fichet:2013ola}). The cutoff of the 4D EFT is $O(\mu)$ above which it is UV-completed by the full braneworld model giving rise to  Eq.~\eqref{eq:AAAA}. In a sense, the presence of the IR brane breaks the scale invariance, which makes perfect sense from the CFT viewpoint. 
From Eq.~\eqref{eq:AAAALE}  one can see that the amplitudes with $E<\mu$ are suppressed by a power of $(E/\mu)^4$ as compared to the scale invariant amplitude Eq.~\eqref{eq:AAAA}. 
From the experimental viewpoint this is just a familiar low-energy behaviour: Experiments with energy scale below $\mu$ tend to be disfavored with respect to those at higher energies.

\section{Conclusion}\label{se:conc}

Braneworld effective theories can be either exactly or quasi-localized. 
In this paper we have argued that, at least in the presence of gravity, an exactly localized theory cannot be obtained by taking a limit in a quasilocalized theory. Exact localization via large bulk masses is obstructed, essentially because 5D gravity couples to bulk masses. Even at the level of a zero mode EFT, gravity robustly ensures that the large bulk mass limit cannot be taken. 
Exact localization via large kinetic term is not obstructed, but does not lead to an exactly localized braneworld  because  a tower of matter KK modes remains in the spectrum and always couples to the brane sector via 5D gravity. Moreover for a gauge field such limit cannot even be taken as it would send the cutoff of the theory to zero, either because of the WGC or because of 5D strong coupling.  

\begin{figure}[t]
	\centering
	\includegraphics[width=0.55\linewidth,trim={1cm 13cm 1cm 3cm},clip]{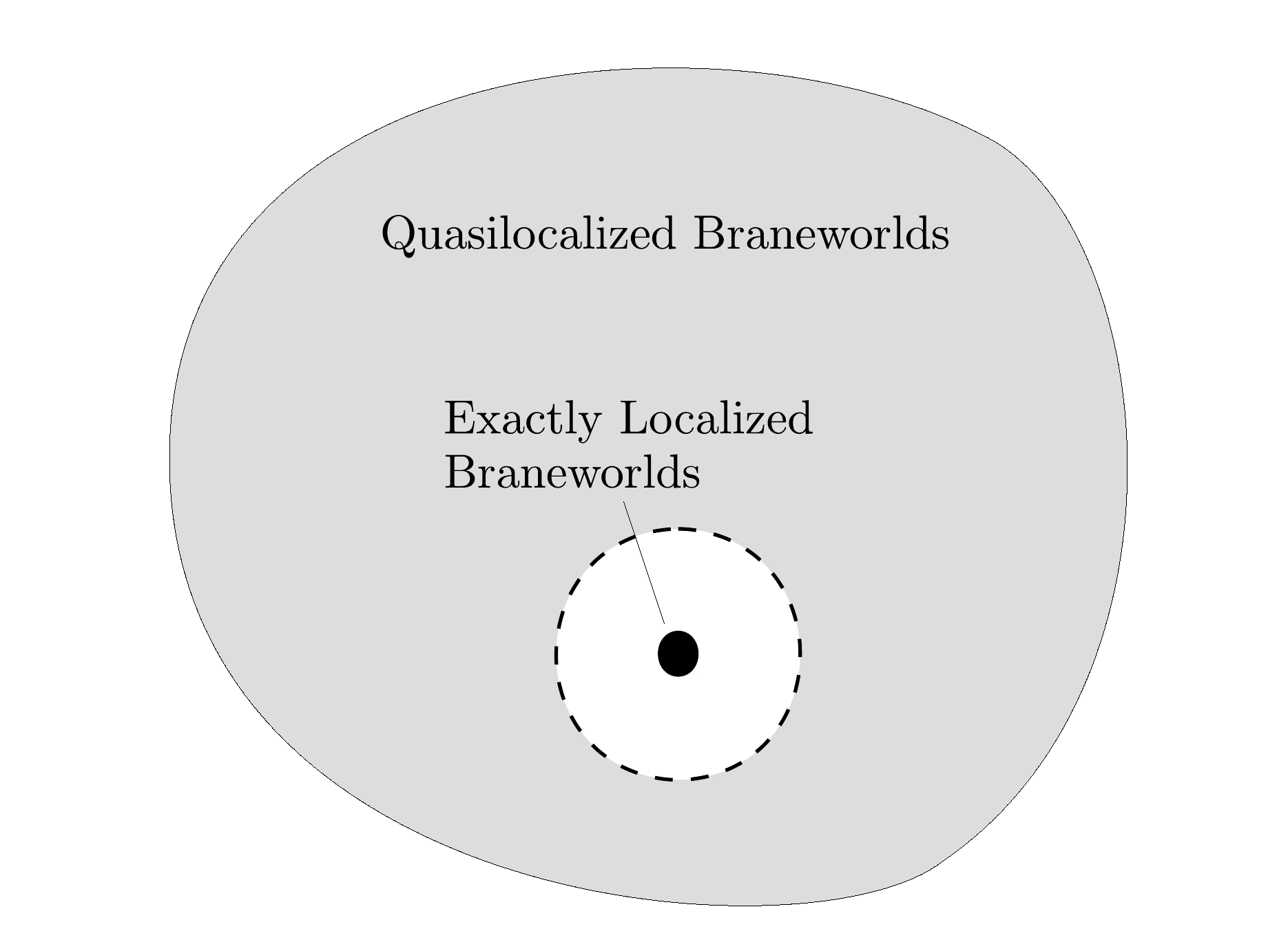}
	\caption{Cartoon of the space of braneworld EFTs with gravity. The gray region represents the parameter space of  quasilocalized braneworld theories. 
		}
	\label{fig:sketch_space}
\end{figure}

Focusing on  exactly localized braneworld EFT, we have presented two simple models in which inconsistencies appear. In a braneworld model with exactly localized matter and a bulk gauge field, we show that an exact global symmetry can exist in the theory, in strict contradiction with expectations from quantum gravity. In a warped model with two branes, inconsistencies appear when the IR brane carries a large number of species.
In both of these models, the paradoxes are solved once the brane fields are made quasilocalized instead of exactly localized. The status of exact vs quasi-localization
in the context of string UV-completions being unclear---at least to us, we do not attempt a generalized claim from the hints obtained on the EFT side.

In any case, all these observations provide excellent motivation to revisit  exactly-localized braneworld embeddings of the SM and make them quasilocalized.  As a general rule, quasilocalization renders the phenomenology of these models richer. 
This is because in quasilocalized models each brane degree of freedom is accompanied by a tower of KK modes---which may possibly be heavy, or may couple to brane fields only via 5D gravity. Additionally, the brane degrees of freedom may have a non-vanishing component of their wavefunction in the bulk, which puts them in direct contact with bulk degrees of freedom.
This bulk component is strictly nonzero for gauge fields. Effects in the gauge sector are quite model-independent as a result of 5D gauge symmetry.  

We focus on the gauge-gravity sector of the quasilocalized warped braneworld. We point out that SM gauge fields have a tree-level anomalous running as a result of the gauge KK modes. The only direction to reduce this effect is to increase the strength of bulk gauge interactions, thereby decreasing the cutoff of the theory. 
We also evaluate the anomalous four-gauge boson scattering induced by 5D gravity. In the pure AdS regime we find that this effect is scale invariant. 
It can thus be probed democratically by experiments at vastly different order of magnitude, which should imply an interesting experimental interplay.

These results from the gauge sector explicitly show that new, somewhat exotic signatures arise from the quasilocalized warped braneworld. Because of AdS/CFT, these effects are reminiscent of those from a conformal hidden sector (see \cite{Brax:2019koq,Costantino:2019ixl} for related dark sector model-building). 
These effects provide new ways to experimentally test and constrain the   hypothesis of the SM being (quasi)localized on a 3-brane.
It would  certainly be interesting to study the other sectors of the quasilocalized warped braneworld.

\section*{Acknowledgments}

I thank    P. Saraswat,  P. Brax, S. Melville, M. Quiros,  G. von Gersdorff and T. Gherghetta  for useful discussions. %
  The author is supported by the S\~ao Paulo Research Foundation (FAPESP) under grants \#2011/11973, \#2014/21477-2 and \#2018/11721-4, and funded in part by the Gordon and Betty Moore Foundation through a Fundamental Physics Innovation Visitor Award (Grant GBMF6210).

\appendix

\section{Propagator in general warped background \label{se:propa_gen}}

Here we show how to compute the scalar propagator in an arbitrarily warped 5D metric (Eq.~\eqref{eq:metric}) and with arbitrary boundary conditions. Our derivation is a standard ODE solving with no shortcut. It is longer than  customary derivations seen in the literature, but it shows explicitly how the structure of the Green function arises. For example our method clarifies the role of the Wronskian, which is not so transparent in the formalism of \cite{Ponton:2012bi}.

Here we write the metric as \be ds^2=g_{MN}dX^MdX^N=\frac{1}{\rho^2(z)}\left(\eta_{\mu\nu}dx^\mu dx^\nu - dz^2 \right)\,.\ee 
We assume $z$ is restricted to an interval  $z\in[a,b]$. 
Consider the scalar equation of motion in the presence of a source,
\be
\partial_M \left( g^{MN} \sqrt{g} \partial_N \Phi \right) +\sqrt{g} M^2 \Phi = {\cal J}(X) \,.
\ee
The Feynman Green function in curved space is defined by 
\be {\cal J}(X)=-i\delta^{(5)}(X-X')\,.\ee

Boundary conditions are assumed to take the generic form
\begin{align}
&{\cal B}_a \Phi\equiv(\alpha_a \partial_5 +\beta_a)\Phi|_{z=a}=0 \,. \nonumber \\
&{\cal B}_b \Phi\equiv(\alpha_b \partial_5 +\beta_b)\Phi|_{z=b}=0 \,. \label{eq:BCs_def}
\end{align}
Introducing $\Phi_p(z)=\int d^4x e^{ip^\mu x_\mu }\Phi(X)$,  the 5D equation of motion becomes
\be
-\partial_5 \left( \rho^{-3} \partial_5 \Phi_p\right) +\left(\rho^{-5} M^2 - \rho^{-3}p^2\right) \Phi_p =  {\cal J}_p(z)\,,
\ee
with $p^2 = \eta_{\mu\nu}p^\mu p^\nu$, or
\be
\partial^2_5 \Phi_p-3\partial_5(\log \rho)\,\, \partial_5 \Phi_p -\left(\rho^{-2} M^2 - p^2\right) \Phi_p = -\rho^3 {\cal J}_p(z)\,.
\ee

The solutions of the homogeneous equation take the form
\be
\Phi_p(z)= A f_p(z)+ B g_p(z)\,, \label{eq:sols}
\ee
where $A$, $B$ are constants.  The Wronskian is defined as \be W(z)=f g'-f' g\,. \label{eq:W_def} \ee Taking its derivative and using the homogeneous equation of motion leads to
\be
W(z)= e^{- \int dz\left( -3 \partial_5(\log \rho)\right)}=C \rho^3\,,
\ee
where $C$ is a  constant. We see that the $\rho$ dependence of $W$ is automatically fixed, only $C$ depends on the solutions Eq.~\eqref{eq:sols}.

The solution to the sourced equation of motion takes the form
\be
\Phi^{\cal J}_p(z)= A(z) f_p(z)+ B(z) g_p(z)\,. 
\ee
Following  standard ODE solving methods, one chooses the condition $A'(z)f(z)+B'(z)g(z)=0  $ and obtains
\begin{align}
& A'(z)=- \frac{g(z)}{W(z)}(-\rho^3) {\cal J}_p(z)= \frac{g(z)}{C} {\cal J}_p(z)\,, \\ & B'(z)= \frac{f(z)}{W(z)}(-\rho^3) {\cal J}_p(z)=-\frac{f(z)}{C} {\cal J}_p(z)\,.
\end{align}
Interestingly, we see that the $\rho$-dependence of the Wronskian always cancels with the $\rho^3$ factor multiplying the source. 

The boundary conditions on $\Phi^{\cal J}_p(z)$ obtained by substituting the general solution Eq.~\eqref{eq:sols} in Eqs~\eqref{eq:BCs_def} 
are 
\begin{align}
&{\cal B}_a \Phi^{\cal J}= (\alpha_a f'(a) +\beta_a f(a)) A(a)+(\alpha_a g'(a) +\beta_a g(a)) B(a)=0 \\
&{\cal B}_b \Phi^{\cal J}= (\alpha_b f'(b) +\beta_b f(b)) A(b)+(\alpha_b g'(b) +\beta_b g(b)) B(b)=0
\end{align}
and one introduces
\begin{align}
&\gamma_a=\alpha_a f'(a) +\beta_a f(a) \,,\\
& \eta_a=\alpha_a g'(a) +\beta_a g(a) \,,\\
& \gamma_b=\alpha_b f'(b) +\beta_b f(b) \,,\\
& \eta_b=\alpha_b g'(b) +\beta_b g(b) \,,
\end{align}
giving
\begin{align}
&{\cal B}_a \Phi^{\cal J}= \gamma_a A(a)+\eta_a B(a)=0 \\
&{\cal B}_b \Phi^{\cal J}= \gamma_b A(b)+\eta_b B(b)=0 \,.
\end{align}
 
 We now see that, in order to obtain $A(z)$, $B(z)$, we can evaluate the integrals of appropriate linear combinations of $A'(z)$, $B'(z)$. We obtain
 \begin{align}
 \int_a^z dz' \left(
 \gamma_a A'(z')+\eta_a B'(z')
 \right)= \gamma_a A(z)+ \eta_a B(z)=  \int_a^zdz' \left(
\gamma_a g(z') -\eta_a f(z')   
 \right)\frac{{\cal J}_p(z')}{C} \\
  \int_z^b dz' \left(
 \gamma_b A'(z')+\eta_b B'(z')
 \right)= - \gamma_b A(z) - \eta_b B(z)=  \int_z^bdz' \left(
\gamma_b g(z') -\eta_b f(z')   
 \right)\frac{{\cal J}_p(z')}{C}
 \end{align}

These relations are conveniently put as a matrix
\be
\begin{pmatrix}
\gamma_a & \eta_a \\
\gamma_b & \eta_b 
\end{pmatrix} 
\begin{pmatrix}
A(z) \\ B(z)
\end{pmatrix}=
\begin{pmatrix}
\int_a^z dz' \left(
\gamma_a g(z') -\eta_a f(z')   
 \right)\frac{{\cal J}_p(z')}{C} \\
  \int_z^b dz' \left(
-\gamma_b g(z') +\eta_b f(z')   
 \right)\frac{{\cal J}_p(z')}{C}
\end{pmatrix}\,.
\ee
Inverting the matrix and replacing $A(z)$, $B(z)$ in  Eq~\eqref{eq:sols}, we obtain the sourced solution
\begin{align}
&\Phi^{\cal J}_p(z)= A(z) f_p(z)+ B(z) g_p(z)=\frac{-1}{\gamma_a \eta_b-\gamma_b \eta_a} \times\\
&
\bigg(\quad\int_a^z dz' \left(
\gamma_b g(z)-\eta_b f(z)  \right) \left(
\gamma_a g(z') -\eta_a f(z')   
 \right)\frac{{\cal J}_p(z')}{C}\quad+ \\
 & \quad \int_z^b dz' \left(
\gamma_a g(z)-\eta_a f(z)  \right) \left(
\gamma_b g(z') -\eta_b f(z')   
 \right)\frac{{\cal J}_p(z')}{C}
\bigg)
\,.
\end{align}
Then observe that this solution can be rewritten as
\be
\Phi^{\cal J}_p(z)= i \int_a^b dz'\, \Delta(z,z'){\cal J}_p(z') \,.
\ee
The $\Delta(z,z')$ is given by
\be
\Delta(z,z')=\frac{i}{C}\frac{\left(
\gamma_a g(z_<) -\eta_a f(z_<)   
 \right)\left(
\gamma_b g(z_>)-\eta_b f(z_>)  \right) }{\gamma_a \eta_b-\gamma_b \eta_a} \, \label{eq:propa_gen}
\ee
which is  the general Feynman propagator for arbitrary boundary conditions and metric.  

In AdS  we have
\be
f(z)=z^2 J_\alpha( p z) \,,\quad  g(z)=z^2 Y_\alpha( p z)
\ee
and obtain
 \be
\rho(z)=k z \,,\quad \sqrt g = (kz)^{-5} \,,\quad W(z)= \frac{2 \,z^3}{\pi} \,,\quad C= \frac{2 }{\pi\,k^3}
\ee
This gives the correct expression
\be
\Delta_{\rm AdS}(z,z')=i\frac{\pi}{2\,k} (kz_>)^2(kz_<)^2\frac{\left(
\tilde J_a Y(pz_<)  -\tilde Y_a J(pz_<)   
 \right)\left(
\tilde J_b Y(pz_>)-\tilde Y_b J(pz_>)  \right) }{\tilde J_a \tilde Y_b-\tilde J_b \tilde Y_a}\,. \label{eq:propa_gen_AdS}
\ee

\bibliographystyle{JHEP}

\bibliography{biblio}

\providecommand{\href}[2]{#2}\begingroup\raggedright\begin{thebibliography}{10}

\bibitem{Polchinski:1996na}
J.~Polchinski, {\it {Tasi lectures on D-branes}},  in {\em {Fields, strings and
  duality. Proceedings, Summer School, Theoretical Advanced Study Institute in
  Elementary Particle Physics, TASI'96, Boulder, USA, June 2-28, 1996}},
  pp.~293--356, 1996.
\newblock \href{http://arxiv.org/abs/hep-th/9611050}{{\tt hep-th/9611050}}.

\bibitem{Bachas:1998rg}
C.~P. Bachas, {\it {Lectures on D-branes}},  in {\em {Duality and
  supersymmetric theories. Proceedings, Easter School, Newton Institute,
  Euroconference, Cambridge, UK, April 7-18, 1997}}, pp.~414--473, 1998.
\newblock \href{http://arxiv.org/abs/hep-th/9806199}{{\tt hep-th/9806199}}.

\bibitem{Aharony:1999ti}
O.~Aharony, S.~S. Gubser, J.~M. Maldacena, H.~Ooguri, and Y.~Oz, {\it {Large N
  field theories, string theory and gravity}},  {\em Phys. Rept.} {\bf 323}
  (2000) 183--386, [\href{http://arxiv.org/abs/hep-th/9905111}{{\tt
  hep-th/9905111}}].

\bibitem{Duff:1996zn}
M.~J. Duff, {\it {Supermembranes}},  in {\em {26th British Universities Summer
  School in Theoretical Elementary Particle Physics (BUSSTEPP 1996) Swansea,
  Wales, September 3-18, 1996}}, 1996.
\newblock \href{http://arxiv.org/abs/hep-th/9611203}{{\tt hep-th/9611203}}.

\bibitem{Csaki:2004ay}
C.~Csaki, {\it {TASI lectures on extra dimensions and branes}},  in {\em {From
  fields to strings: Circumnavigating theoretical physics. Ian Kogan memorial
  collection (3 volume set)}}, pp.~605--698, 2004.
\newblock \href{http://arxiv.org/abs/hep-ph/0404096}{{\tt hep-ph/0404096}}.
\newblock [,967(2004)].

\bibitem{Sundrum:1998sj}
R.~Sundrum, {\it {Effective field theory for a three-brane universe}},  {\em
  Phys. Rev.} {\bf D59} (1999) 085009,
  [\href{http://arxiv.org/abs/hep-ph/9805471}{{\tt hep-ph/9805471}}].

\bibitem{Sundrum:1998ns}
R.~Sundrum, {\it {Compactification for a three-brane universe}},  {\em Phys.
  Rev.} {\bf D59} (1999) 085010,
  [\href{http://arxiv.org/abs/hep-ph/9807348}{{\tt hep-ph/9807348}}].

\bibitem{Dvali:2000hr}
G.~R. Dvali, G.~Gabadadze, and M.~Porrati, {\it {4-D gravity on a brane in 5-D
  Minkowski space}},  {\em Phys. Lett.} {\bf B485} (2000) 208--214,
  [\href{http://arxiv.org/abs/hep-th/0005016}{{\tt hep-th/0005016}}].

\bibitem{Randall:1999ee}
L.~Randall and R.~Sundrum, {\it {A Large mass hierarchy from a small extra
  dimension}},  {\em Phys. Rev. Lett.} {\bf 83} (1999) 3370--3373,
  [\href{http://arxiv.org/abs/hep-ph/9905221}{{\tt hep-ph/9905221}}].

\bibitem{Randall:1999vf}
L.~Randall and R.~Sundrum, {\it {An Alternative to compactification}},  {\em
  Phys. Rev. Lett.} {\bf 83} (1999) 4690--4693,
  [\href{http://arxiv.org/abs/hep-th/9906064}{{\tt hep-th/9906064}}].

\bibitem{Brax:2019koq}
P.~Brax, S.~Fichet, and P.~Tanedo, {\it {The Warped Dark Sector}},  {\em Phys.
  Lett.} {\bf B798} (2019) 135012, [\href{http://arxiv.org/abs/1906.0219}{{\tt
  arXiv:1906.0219}}].

\bibitem{Costantino:2019ixl}
A.~Costantino, S.~Fichet, and P.~Tanedo, {\it {Exotic Spin-Dependent Forces
  from a Hidden Sector}},  \href{http://arxiv.org/abs/1910.0297}{{\tt
  arXiv:1910.0297}}.

\bibitem{Nastase:2007kj}
H.~Nastase, {\it {Introduction to AdS-CFT}},
  \href{http://arxiv.org/abs/0712.0689}{{\tt arXiv:0712.0689}}.

\bibitem{Gherghetta:2010cj}
T.~Gherghetta, {\it {A Holographic View of Beyond the Standard Model Physics}},
   in {\em {Physics of the large and the small, TASI 09, proceedings of the
  Theoretical Advanced Study Institute in Elementary Particle Physics, Boulder,
  Colorado, USA, 1-26 June 2009}}, pp.~165--232, 2011.
\newblock \href{http://arxiv.org/abs/1008.2570}{{\tt arXiv:1008.2570}}.

\bibitem{Ponton:2012bi}
E.~Ponton, {\it {TASI 2011: Four Lectures on TeV Scale Extra Dimensions}},  in
  {\em {The Dark Secrets of the Terascale: Proceedings, TASI 2011, Boulder,
  Colorado, USA, Jun 6 - Jul 11, 2011}}, pp.~283--374, 2013.
\newblock \href{http://arxiv.org/abs/1207.3827}{{\tt arXiv:1207.3827}}.

\bibitem{Witten:1998qj}
E.~Witten, {\it {Anti-de Sitter space and holography}},  {\em Adv. Theor. Math.
  Phys.} {\bf 2} (1998) 253--291,
  [\href{http://arxiv.org/abs/hep-th/9802150}{{\tt hep-th/9802150}}].

\bibitem{Batell:2007jv}
B.~Batell and T.~Gherghetta, {\it {Holographic mixing quantified}},  {\em Phys.
  Rev.} {\bf D76} (2007) 045017, [\href{http://arxiv.org/abs/0706.0890}{{\tt
  arXiv:0706.0890}}].

\bibitem{Cabrer:2009we}
J.~A. Cabrer, G.~von Gersdorff, and M.~Quiros, {\it {Soft-Wall Stabilization}},
   {\em New J. Phys.} {\bf 12} (2010) 075012,
  [\href{http://arxiv.org/abs/0907.5361}{{\tt arXiv:0907.5361}}].

\bibitem{Fitzpatrick:2010zm}
A.~L. Fitzpatrick, E.~Katz, D.~Poland, and D.~Simmons-Duffin, {\it {Effective
  Conformal Theory and the Flat-Space Limit of AdS}},  {\em JHEP} {\bf 07}
  (2011) 023, [\href{http://arxiv.org/abs/1007.2412}{{\tt arXiv:1007.2412}}].

\bibitem{Dudas:2012mv}
E.~Dudas and G.~von Gersdorff, {\it {Universal contributions to scalar masses
  from five dimensional supergravity}},  {\em JHEP} {\bf 10} (2012) 100,
  [\href{http://arxiv.org/abs/1207.0815}{{\tt arXiv:1207.0815}}].

\bibitem{Carena:2002dz}
M.~Carena, E.~Ponton, T.~M.~P. Tait, and C.~E.~M. Wagner, {\it {Opaque Branes
  in Warped Backgrounds}},  {\em Phys. Rev.} {\bf D67} (2003) 096006,
  [\href{http://arxiv.org/abs/hep-ph/0212307}{{\tt hep-ph/0212307}}].

\bibitem{ArkaniHamed:2006dz}
N.~Arkani-Hamed, L.~Motl, A.~Nicolis, and C.~Vafa, {\it {The String landscape,
  black holes and gravity as the weakest force}},  {\em JHEP} {\bf 06} (2007)
  060, [\href{http://arxiv.org/abs/hep-th/0601001}{{\tt hep-th/0601001}}].

\bibitem{Rubakov:1983bb}
V.~A. Rubakov and M.~E. Shaposhnikov, {\it {Do We Live Inside a Domain Wall?}},
   {\em Phys. Lett.} {\bf 125B} (1983) 136--138.

\bibitem{ArkaniHamed:1999za}
N.~Arkani-Hamed, Y.~Grossman, and M.~Schmaltz, {\it {Split fermions in extra
  dimensions and exponentially small cross-sections at future colliders}},
  {\em Phys. Rev.} {\bf D61} (2000) 115004,
  [\href{http://arxiv.org/abs/hep-ph/9909411}{{\tt hep-ph/9909411}}].

\bibitem{Mirabelli:1999ks}
E.~A. Mirabelli and M.~Schmaltz, {\it {Yukawa hierarchies from split fermions
  in extra dimensions}},  {\em Phys. Rev.} {\bf D61} (2000) 113011,
  [\href{http://arxiv.org/abs/hep-ph/9912265}{{\tt hep-ph/9912265}}].

\bibitem{Palti:2019pca}
E.~Palti, {\it {The Swampland: Introduction and Review}},  2019.
\newblock \href{http://arxiv.org/abs/1903.0623}{{\tt arXiv:1903.0623}}.

\bibitem{Fichet:2019ugl}
S.~Fichet and P.~Saraswat, {\it {Approximate Symmetries and Gravity}},  {\em
  {\rm to appear in} JHEP} [\href{http://arxiv.org/abs/1909.0200}{{\tt
  arXiv:1909.0200}}].

\bibitem{ArkaniHamed:2000ds}
N.~Arkani-Hamed, M.~Porrati, and L.~Randall, {\it {Holography and
  phenomenology}},  {\em JHEP} {\bf 08} (2001) 017,
  [\href{http://arxiv.org/abs/hep-th/0012148}{{\tt hep-th/0012148}}].

\bibitem{Polchinski:2002jw}
J.~Polchinski and M.~J. Strassler, {\it {Deep inelastic scattering and gauge /
  string duality}},  {\em JHEP} {\bf 05} (2003) 012,
  [\href{http://arxiv.org/abs/hep-th/0209211}{{\tt hep-th/0209211}}].

\bibitem{Gherghetta:2003he}
T.~Gherghetta and A.~Pomarol, {\it {The Standard model partly supersymmetric}},
   {\em Phys. Rev.} {\bf D67} (2003) 085018,
  [\href{http://arxiv.org/abs/hep-ph/0302001}{{\tt hep-ph/0302001}}].

\bibitem{Fichet:2019hkg}
S.~Fichet, {\it {Opacity and effective field theory in anti–de Sitter
  backgrounds}},  {\em Phys. Rev.} {\bf D100} (2019), no.~9 095002,
  [\href{http://arxiv.org/abs/1905.0577}{{\tt arXiv:1905.0577}}].

\bibitem{Dvali:2007hz}
G.~Dvali, {\it {Black Holes and Large N Species Solution to the Hierarchy
  Problem}},  {\em Fortsch. Phys.} {\bf 58} (2010) 528--536,
  [\href{http://arxiv.org/abs/0706.2050}{{\tt arXiv:0706.2050}}].

\bibitem{Freedman:1999gp}
D.~Z. Freedman, S.~S. Gubser, K.~Pilch, and N.~P. Warner, {\it {Renormalization
  group flows from holography supersymmetry and a c theorem}},  {\em Adv.
  Theor. Math. Phys.} {\bf 3} (1999) 363--417,
  [\href{http://arxiv.org/abs/hep-th/9904017}{{\tt hep-th/9904017}}].

\bibitem{Myers:2010tj}
R.~C. Myers and A.~Sinha, {\it {Holographic c-theorems in arbitrary
  dimensions}},  {\em JHEP} {\bf 01} (2011) 125,
  [\href{http://arxiv.org/abs/1011.5819}{{\tt arXiv:1011.5819}}].

\bibitem{Aldazabal:2000sa}
G.~Aldazabal, L.~E. Ibanez, F.~Quevedo, and A.~M. Uranga, {\it {D-branes at
  singularities: A Bottom up approach to the string embedding of the standard
  model}},  {\em JHEP} {\bf 08} (2000) 002,
  [\href{http://arxiv.org/abs/hep-th/0005067}{{\tt hep-th/0005067}}].

\bibitem{Antoniadis:2000jv}
I.~Antoniadis, K.~Benakli, and A.~Laugier, {\it {Contact interactions in
  D-brane models}},  {\em JHEP} {\bf 05} (2001) 044,
  [\href{http://arxiv.org/abs/hep-th/0011281}{{\tt hep-th/0011281}}].

\bibitem{Uranga:933469}
A.~M. Uranga, {\it {TASI lectures on string compactification, model building
  and fluxes}},  Tech. Rep. CERN-PH-TH-2005-205. IFT-UAM-CSIC-05-044, CERN,
  Geneva, 2005.

\bibitem{Moeller:2000jy}
N.~Moeller, A.~Sen, and B.~Zwiebach, {\it {D-branes as tachyon lumps in string
  field theory}},  {\em JHEP} {\bf 08} (2000) 039,
  [\href{http://arxiv.org/abs/hep-th/0005036}{{\tt hep-th/0005036}}].

\bibitem{Randall:2001gb}
L.~Randall and M.~D. Schwartz, {\it {Quantum field theory and unification in
  AdS5}},  {\em JHEP} {\bf 11} (2001) 003,
  [\href{http://arxiv.org/abs/hep-th/0108114}{{\tt hep-th/0108114}}].

\bibitem{Chacko:1999hg}
Z.~Chacko, M.~A. Luty, and E.~Ponton, {\it {Massive higher dimensional gauge
  fields as messengers of supersymmetry breaking}},  {\em JHEP} {\bf 07} (2000)
  036, [\href{http://arxiv.org/abs/hep-ph/9909248}{{\tt hep-ph/9909248}}].

\bibitem{Agashe:2007zd}
K.~Agashe, H.~Davoudiasl, G.~Perez, and A.~Soni, {\it {Warped Gravitons at the
  LHC and Beyond}},  {\em Phys. Rev.} {\bf D76} (2007) 036006,
  [\href{http://arxiv.org/abs/hep-ph/0701186}{{\tt hep-ph/0701186}}].

\bibitem{Goldberger:2002cz}
W.~D. Goldberger and I.~Z. Rothstein, {\it {High-energy field theory in
  truncated AdS backgrounds}},  {\em Phys. Rev. Lett.} {\bf 89} (2002) 131601,
  [\href{http://arxiv.org/abs/hep-th/0204160}{{\tt hep-th/0204160}}].

\bibitem{Contino:2002kc}
R.~Contino, P.~Creminelli, and E.~Trincherini, {\it {Holographic evolution of
  gauge couplings}},  {\em JHEP} {\bf 10} (2002) 029,
  [\href{http://arxiv.org/abs/hep-th/0208002}{{\tt hep-th/0208002}}].

\bibitem{Goldberger:2002hb}
W.~D. Goldberger and I.~Z. Rothstein, {\it {Effective field theory and
  unification in AdS backgrounds}},  {\em Phys. Rev.} {\bf D68} (2003) 125011,
  [\href{http://arxiv.org/abs/hep-th/0208060}{{\tt hep-th/0208060}}].

\bibitem{Boos:2002hf}
E.~E. Boos, Y.~A. Kubyshin, M.~N. Smolyakov, and I.~P. Volobuev, {\it
  {Effective Lagrangians for physical degrees of freedom in the Randall-Sundrum
  model}},  {\em Class. Quant. Grav.} {\bf 19} (2002) 4591--4606,
  [\href{http://arxiv.org/abs/hep-th/0202009}{{\tt hep-th/0202009}}].

\bibitem{Hinterbichler:2011tt}
K.~Hinterbichler, {\it {Theoretical Aspects of Massive Gravity}},  {\em Rev.
  Mod. Phys.} {\bf 84} (2012) 671--710,
  [\href{http://arxiv.org/abs/1105.3735}{{\tt arXiv:1105.3735}}].

\bibitem{Fichet:2014uka}
S.~Fichet, G.~von Gersdorff, B.~Lenzi, C.~Royon, and M.~Saimpert, {\it
  {Light-by-light scattering with intact protons at the LHC: from Standard
  Model to New Physics}},  {\em JHEP} {\bf 02} (2015) 165,
  [\href{http://arxiv.org/abs/1411.6629}{{\tt arXiv:1411.6629}}].

\bibitem{Fichet:2013ola}
S.~Fichet and G.~von Gersdorff, {\it {Anomalous gauge couplings from composite
  Higgs and warped extra dimensions}},  {\em JHEP} {\bf 03} (2014) 102,
  [\href{http://arxiv.org/abs/1311.6815}{{\tt arXiv:1311.6815}}].

\end{thebibliography}\endgroup

\end{document}